\documentclass[aps,groupedaddress,nofootinbib]{revtex4-1}
\usepackage{amssymb, amsmath, siunitx}
\usepackage{hyperref}
\usepackage{slashed}
\usepackage{bbold}
\usepackage{graphicx}

\newcommand{\Tr}{\ensuremath{\text{Tr}}}

\begin{document}


\title{Dynamical Supergravity Breaking via the Super-Higgs Effect Revisited}

\author{Jean Alexandre$^a$, Nick Houston$^a$ and Nick E. Mavromatos$^{a,b}$}

\affiliation{$^a$Theoretical Particle Physics and Cosmology Group, Physics Department, King's College London, Strand, London WC2R 2LS. \\
 $^b$ Also currently at:  Theory Division, Physics Department, CERN, CH-1211 Geneva 23, Switzerland.}


\begin{flushleft}
KCL-PH-TH/2013-\textbf{21} \\
LCTS/2013-12
\end{flushleft} 

\begin{abstract} 
\vspace{0.2cm}
We investigate the dynamical breaking of local supersymmetry (supergravity), including the Deser-Zumino super-Higgs effect, via the corresponding one-loop effective potential for the simple but quite representative cases of $\mathcal{N}=1$, $D=4$ simple supergravity and a (simplified) conformal version of it. We find solutions to the effective equations which indicate dynamical generation of a gravitino mass, thus breaking supergravity. In the case of conformal supergravity models,  the gravitino mass can be much lower than the Planck scale, for global supersymmetry breaking scales below the Grand Unification scale. The absence of instabilities in the effective potential arising from the quantum fluctuations of the metric field is emphasised, contrary to previous claims in the literature.  
\end{abstract}

\maketitle

\tableofcontents

\section{Introduction and Summary} 

Supersymmetry is an important theoretical and phenomenological subject that has dominated theoretical physics
for decades, despite the lack (so far) of evidence that it actually exists in Nature. 
It assists us in understanding several aspects of low energy physics, from the stability of the Higgs vacuum to the unification of forces, and even the low scale of inflation (relative to the Planck mass), in a way that is much clearer than in non-supersymmetric frameworks. 
The embedding of supersymmetry into a gravitational framework by promoting global supersymmetry into a local (gauged) symmetry, termed supergravity~\cite{VanNieuwenhuizen:1981ae}, has initiated many interesting avenues of research toward understanding both the unification of gravity with the remaining three fundamental forces of Nature, and to some extent quantisation of the gravitational field.
In particular, it is well known that supergravity theories constitute consistent low-energy limits of superstring theories, which are thought to provide plausible paths to the quantisation of gravity in concert with other fundamental interactions. \\

However, supersymmetry is not phenomenologically observed in Nature, and thus if it exists, it must be broken in the low energy world. 
It is therefore desirable to understand the spontaneous or dynamical breaking of supersymmetry under various possible circumstances, of which, thanks to the profusion of gauge and matter sectors that may be consistently coupled into supersymmetry, there are many \cite{nilles}.
It would be preferable however, to seek a way of breaking local supersymmetry (supergravity) directly in a dynamical manner, without the need to couple it to a gauge sector. \\

One such approach would be to leverage the fermionic (gravitino) torsion terms that are generically present in supergravity theories, which consist of four-gravitino self-interaction terms. 
The latter could conceivably condense under certain circumstances, thereby producing condensates of the gravitino field, which in this way would dynamically acquire a non-zero mass whilst leaving the graviton massless.
In this way supergravity would be broken \emph{dynamically}, in the same spirit as the breaking of chiral symmetry in the Nambu-Jona-Lasinio model~\cite{NJL}. \\

It was conjectured in \cite{smith,smith2} that such a dynamical breaking of supergravity could occur via the formation of condensates of the gravitino field, with analyses based on the one-loop effective potential of a simple ${\mathcal N}=1$ supergravity model in a flat Minkowski space-time (necessary to allow an unambiguous definition of the gravitino mass via the condensate field). 
The effective potential of the gravitino condensate field, dependent on positive powers of the ultraviolet (UV) cutoff scale due to the well-known non-renormalisability of quantum gravity in four space-time dimensions, acquires a non-trivial minimum for some values of the cutoff relative to the gravitational constant (reduced Planck mass).  
In the analyses of \cite{smith,smith2} the gravitino thereby acquires a dynamical mass of the order of the Planck scale. \\

It was important for their analyses that the one-loop value of the effective potential \emph{vanishes} at the non-trivial minimum, which implies the vanishing of the effective vacuum energy of the resulting low-energy theory. 
This arguably justifies \emph{a posteriori} the Minkowski space-time analysis of the effective potential. 
It was also important for both the entire approach and the double-well shape of the effective potential, appropriate for supergravity breaking, that the Deser-Zumino super-Higgs effect~\cite{deser} was incorporated self consistently by coupling the supergravity action with the non-linear Volkov-Akulov action~\cite{va} of the Goldstone particle associated with the (assumed F-type) spontaneous global supersymmetry breaking. 
The latter is a Majorana spin 1/2 fermion, the Goldstino. \\

This formalism was essential for two reasons. 
Firstly, the Goldstino could be absorbed by the gravitino, under an appropriate field redefinition to make the latter massive, and thus disappear from the physical spectrum
Secondly, the only subsequent trace of the Goldstino would be a (negative) cosmological constant $-f^2$ in the effective action, which is associated with the scale of global supersymmetry breaking; namely the vacuum expectation value of the appropriate F-term of the chiral scalar superfield responsible for spontaneous supersymmetry breaking. 
This is the super-Higgs effect~\cite{deser} which is essential in keeping track of the right degrees of freedom in the problem of dynamical acquisition of a mass by the gravitino field, and the consequent breaking of local supersymmetry. \\

From the point of view of the effective potential, this would correspond to a \emph{positive} term at the origin in field space of order of $f^2$, which is responsible for the double-well shape of the potential at the broken symmetry phase. These considerations are consistent with the generic features of dynamical supersymmetry breaking outlined by Witten \cite{witten}, according to which the vacuum energy of broken global supersymmetry is \emph{necessarily positive}, here $f^2 > 0$, whilst a \emph{broken local supersymmetry} (supergravity) can still be characterised by \emph{zero vacuum energy.}\\

Although physically appealing, the flat Minkowski space-time approach of \cite{smith,smith2} was criticised in \cite{Buchbinder:1989gi} due to the fact that it ignored the quantum fluctuations of the metric field. 
Following the generic approach of Fradkin and Tseytlin \cite{Fradkin:1983mq} of calculating the one-loop effective potential in four-dimensional supergravity theories by means of expanding about a generic (anti)de Sitter, rather than Minkowski, background, the authors of \cite{Buchbinder:1989gi} have argued that integrating over metric fluctuations introduces \emph{imaginary} parts in the effective potential, for any non-trivial value of the gravitino condensate field, and irrespective of the value of the background cosmological constant $\Lambda$ (whose role is to effectively replace \cite{Fradkin:1983mq} the flat-space UV cut-off of \cite{smith,smith2}).\\

The presence of imaginary parts would of course be an indication that the non-trivial-gravitino-condensate (broken supergravity) vacuum is \emph{unstable}, and thus there would be no possibility of breaking ${\mathcal N}=1$ supergravity dynamically in this simple and direct way. 
Of course, the traditional way of breaking supergravity via dynamical global supersymmetry breaking through, say, gaugino condensation \cite{nilles}, which would then be communicated to the (super)gravity sector, evades the arguments of \cite{Buchbinder:1989gi} and thus has appeared to be the only consistent way of dynamically breaking supergravity, carrying the price of necessarily coupling the theory to matter fields. \\

In this work we shall revisit the arguments and the analysis of \cite{Buchbinder:1989gi}, with a view to incorporating the super-Higgs effect that was not included in their analyses. 
We have been motivated to do so by the compelling nature and simplicity of the possibility of direct dynamical breaking of supergravity by means of its gravitino-torsion self-interactions. \\

As it turns out, and as is demonstrated below in great detail, the proper incorporation of the super-Higgs effect in such a framework enables the dynamical breaking of supergravity (prior to its coupling to matter or gauge fields), in the sense that a one-loop effective potential analysis considering fully metric fluctuations about (anti)de Sitter backgrounds, and thereby fully incorporating any (weak) quantum gravitational effects, reveals the existence of non-trivial vacua with \emph{no imaginary} parts, contrary to the claims of \cite{Buchbinder:1989gi}. 
The double-well shape of the effective potential, and its vanishing at the non-trivial minima, fully justifies the flat-space approximate analysis of \cite{smith,smith2} and demonstrates that the presence of the cosmological constant $f^2$ due to the super-Higgs effect is responsible for the vanishing of the effective vacuum energy of the low-energy effective action of supergravity at the one-loop level. \\

The structure of the article is as follows.
\begin{itemize}
	\item In section \ref{sec:superhiggs} we briefly review the super-Higgs effect in the context of simple ${\mathcal N}=1, D=4$ supergravity, in order to outline to the reader its important features, and the underlying physics that will be used in our analysis of the resultant one-loop effective potential. 
	\item In section \ref{sec:sugra} we set up the basic formalism and notations underlying the model of ${\mathcal N}=1$ supergravity that we shall consider for concreteness in this work. 
Our formalism of course applies in general to more complicated theories of supergravity. 
	\item In section \ref{sec:quadr} we discuss the quadratic action obtained from previous section by incorporating weak metric fluctuations about a given (anti)de Sitter background, in conjunction with the linearisation of four-gravitino terms by means of appropriate auxiliary fields. 
This is the first step towards the construction of a `proper' (i.e. incorporating (weak) quantum gravitational effects) one-loop effective potential. 
We discuss bosonic and fermionic sectors separately as well as their respective gauge fixing procedures. 
	\item In section \ref{sec:potential} we construct the effective potential and discuss the shape that we require in order to obtain non-trivial local supersymmetry (supergravity) breaking, via dynamical condensation of the gravitino field. 
We discuss carefully the absence of imaginary parts in certain non-trivial vacua, contrary to the claims of \cite{Buchbinder:1989gi}, and explain carefully how this may be achieved. As we also show in subsection \ref{sec:confsugra}, phenomenologically realistic values for the gravitino mass and the associated breaking scale of global supersymmetry can be achieved only in appropriate conformal extensions of the ${\mathcal N}=1$ supergravity, which may also be relevant for inflation.   
	\item Conclusions and outlook are then given in section \ref{sec:concl}, and some technical aspects of our approach of constructing the one-loop effective potential are discussed in an Appendix. 
\end{itemize}

\section{Super-Higgs Effect and Goldstino coupling to supergravity \label{sec:superhiggs}} 

The Goldstino is a Majorana fermion with a Volkov-Akulov lagrangian \cite{va} that arises from some spontaneous or dynamical breaking of global supersymmetry, as a result of the appropriate extension of Goldstone's theorem to supersymmetry
\footnote{The reader is reminded that the generator $Q^\alpha $ of global supersymmetry transformations is a space-time spinor, and hence if $\omega$ denotes the Goldstone field, then the commutator that is not annihilating the vacuum ($\epsilon^\alpha $ is an infinitesimal global-supersymmetry-transformation parameter) $\langle 0 \, | \delta_\varepsilon \omega \, | \, 0  \rangle = -\varepsilon^\alpha \, \langle 0 \, | \, [ {Q}_\alpha \, , \, \omega \, ] \, | \, 0 \rangle - \overline{\varepsilon}^{\dot \alpha} \,   \langle 0 \, | \, [  ({Q}_\alpha)^\dagger \, , \, \omega \, ] \, | \, 0 \rangle \ne 0$ must be a scalar for reasons of Lorentz invariance, and hence $\omega $ must be a fermionic field, to be identified with a spin-1/2 four-component Majorana Goldstone fermion (Goldstino), $\lambda $.}. 
Here we consider the case where the breaking of global supersymmetry is of the so-called $F$-type, in which the $F$-term of some chiral superfield $\Phi$ acquires a vacuum expectation value 
	\begin{equation}\label{fterm}
		\langle F \rangle = f ~.
	\end{equation}
In the original works of~\cite{va}, the Goldstino lagrangian is written in terms of a four-component Majorana spin-$1/2$ Goldstino field (denoted $\lambda $), with $\mathcal{L}_\lambda = -(f^2){\det}\left(\delta^\mu_\nu + i\overline{\lambda} \gamma^\nu \partial_\mu \lambda/2f^2 \right)$. 
The constant $f$ expresses the strength of global supersymmetry breaking, which as mentioned above occurs in our case at $f=1/2$ in Planck units, and the
lagrangian is characterised by a non-linear realisation of global supersymmetry with infinitesimal parameter $\alpha$:
\begin{equation}\label{trnl}
\delta \lambda = f\,\alpha + i \frac{1}{f} \overline{\alpha} \gamma^\mu \lambda \partial_\mu \lambda~.
\end{equation}
The coupling of the Goldstino to supergravity may then generate a mass for the gravitino through the absorption of the Goldstino, via the super-Higgs effect envisaged in \cite{deser}.\\

According to this model, the $\mathcal{N}=1$ supergravity theory is coupled to the Goldstino field $\lambda$. 
Thus, we consider a spontaneously-broken supersymmetric theory with a Majorana Goldstino $\lambda$,
whose action takes the non-linear form considered by Volkov and Akulov~\cite{va,deser}:
	\begin{eqnarray}\label{goldstino}
		\mathcal{L}_\lambda = -f^2 {\rm det}\left(\delta^\mu_\nu + i\, \frac{1}{2 f^2} \, \overline{\lambda} \gamma^\mu \partial_\mu \lambda \right) = - f^2 - \frac{1}{2} i \overline{\lambda} \gamma^\mu \partial_\mu \lambda + \dots~,
	\end{eqnarray}
where we used a weak field expansion expansion of the determinant~\footnote{It is worth noticing that, on using a two-component (Weyl) representation of the Majorana Goldstino field, $\lambda = \begin{pmatrix} G^c \\ G \end{pmatrix} $,
where 
 $G$ is a two-component Weyl spinor,  employing fermionic spinor truncation, the  Volkov-Akulov Lagrangian acquires the exact form:
	\begin{equation}\label{va3b}
		\mathcal{L}_{\rm VA-2\, comp} = -f^2 + i \partial_\mu {\overline G} \overline{\sigma}^\mu G + \frac{1}{2f^2}\, \overline{G}^2 \partial^2 G^2 - 		\frac{1}{4\, f^4}\, G^2 {\overline G}^2 \partial^2 G^2 \partial^2 {\overline G}^2~. \nonumber
	\end{equation} 
For our purposes here, truncation to terms of first order in derivatives suffices in the weak field approximation we utilise herein.}.
Here we keep the discussion general by allowing for an arbitrary value of the parameter $f$.\\

As discussed in~\cite{deser}, one can promote the global supersymmetry to a local one, by allowing the parameter $\alpha \left(x\right)$ to depend on space-time coordinates, and coupling the action (\ref{goldstino}) to that of $\mathcal{N}=1$ supergravity in such a way that the combined action is invariant under the following supergravity transformations:
	\begin{eqnarray}\label{sugratransgolds}
		\delta \lambda &=& f\, \alpha\left(x\right) + \dots ~, \nonumber \\
		\delta e^a _\mu  & = & -i \kappa \overline{\alpha}\left(x\right) \gamma^a \psi_\mu~, \nonumber \\
		\delta \psi_\mu & = & - 2 \kappa^{-1} \partial_\mu \alpha\left(x\right) + \dots
	\end{eqnarray}
where the $\dots $ in the $\lambda$ transformation denote non-linear $\lambda$-dependent terms (\emph{cf.} the variation \eqref{trnl}). 
The action that changes by a divergence under these transformations is the standard $\mathcal{N}=1$ supergravity action plus
	\begin{equation}\label{va2b}
		\mathcal{L}_\lambda = - f^2 e - \frac{i}{2}\overline{\lambda} \gamma^\mu \partial_\mu \lambda - \frac{i\,f}{\sqrt{2}} \overline{\lambda} \gamma^\nu \psi_\nu + \dots ,
	\end{equation}
which contains the coupling of the Goldstino to the gravitino. 
In the above notation, $e$ denotes the vierbein determinant.\\

The Goldstino can then be gauged away \cite{deser} by a suitable redefinition of the gravitino field and the tetrad.
One may impose the gauge condition
	\begin{equation}\label{gravinogauge}
		\psi_\mu \gamma^\mu = 0~,
	\end{equation}
but this leaves behind a \emph{negative cosmological constant term}, $-f^2\, e$, so the total Lagrangian after these redefinitions reads:
	\begin{equation}\label{va3}
		\mathcal{L}_{\rm eff} = -f^2e + (\mathcal{N}=1~{\rm supergravity})~, 
	\end{equation}
where the lagrangian of $\mathcal{N}=1$ supergravity is the subject of the next section. \\

\section{$\mathcal{N}=1$, $D=4$ simple supergravity: preliminaries \label{sec:sugra}}

Working in the 1.5 order formalism of the Einstein-Cartan formulation of supergravity, we have the lagrangian \cite{VanNieuwenhuizen:1981ae}
	\begin{align}\label{lagrangian}
		&\mathcal{L}_{SG}=-\frac{e}{2\kappa^2}R(e,\psi)
		-\frac{1}{2}\epsilon^{\mu\nu\lambda\rho}\overline{\psi_\mu}\gamma_5\gamma_\nu D_\lambda\psi_\rho
		+\frac{e}{3}\left(A_\mu^2-S^2-P^2\right)~,\quad 
		\kappa^2=8\pi G~,\\
		&R\left(e,\psi\right)
		=e^\mu{}_a e^\nu{}_b R^{ab}{}_{\mu\nu}\left(e,\psi\right)
		=R\left(e\right)+\frac{11}{4}\kappa^4\left(\overline{\psi_\mu}\Gamma^{\mu\nu}\psi_\nu\right)^2+\dots,\quad
		\Gamma^{\mu\nu}=\frac{1}{4}\gamma^{[\mu}\gamma^{\nu]}~,\label{Rexpansion}\\
		&R^{ab}{}_{\mu\nu}\left(e,\psi\right)
		=\partial_\mu\omega^{ab}{}_\nu\left(e,\psi\right)
		-\partial_\nu\omega^{ab}{}_\mu\left(e,\psi\right)
		+\omega_\mu{}^{ac}\omega_{\nu c}{}^b\left(e,\psi\right)
		-\omega_\nu{}^{ac}\omega_{\mu c}{}^b\left(e,\psi\right)~,\\
		&\omega_\mu{}^{ab}\left(e,\psi\right)
		=\omega_\mu{}^{ab}\left(e\right)
		+\frac{\kappa^2}{4}\left(\overline\psi_\mu\gamma^a \psi^b
		-\overline\psi_\mu\gamma^b \psi^a
		+\overline\psi^a\gamma_\mu \psi^b\right)~,\quad
		D_\mu=\partial_\mu+\frac{1}{2}\omega_\mu{}^{ab}\left(e\right)\sigma_{ab}~,
	\end{align}	
where $R\left[e\right]$ is the scalar curvature in the absence of torsion, $\left(A_\mu,\;S,\;P\right)$ are the minimal set of auxiliary fields required for closure of the algebra and $\dots$ indicates interaction terms between the gravitino and graviton fields, and four-gravitino interactions involving $\gamma_5$.
As we are in the 1.5 order formalism our spin connection $\omega\left(e\right)$ is determined entirely by the associated field equation, so that we may neglect the variation $\delta\omega$.\\

The action is also invariant under the local supersymmetry transformations
	\begin{align}\label{susy transformations}
		&\delta e^a{}_\mu=\frac{\kappa}{2}\overline{\epsilon}\;\gamma^a\psi_\mu~, \quad
		\delta \psi_\mu=\frac{1}{\kappa}\left(D_\mu+\frac{i\kappa}{2}A_\mu\gamma^5\right)\epsilon-\frac{1}{2}\gamma_\mu\eta\;\epsilon~, \quad
		\delta S=\frac{1}{4}\overline{\epsilon}\;\gamma\cdot \overline{R}~, \\\nonumber
		&\delta P=-\frac{i}{4}\overline{\epsilon}\;\gamma_5\gamma\cdot \overline{R}~,\quad
		\delta A_\mu=\frac{3i}{4}\overline{\epsilon}\;\gamma_5\left(\overline{R}_\mu-\frac{1}{3}\gamma_\mu\gamma\cdot \overline{R}\right)~, \quad
		\eta=-\frac{1}{3}\left(S-i\gamma_5 P-i\slashed{A}\gamma_5\right)~,
	\end{align}
where $\overline{R}$ is the `supercovariantised' gravitino field equation
	\begin{align}
		\overline{R}^{\mu}=\epsilon^{\mu\nu\rho\sigma}\gamma_5\gamma_\nu\left(D_\rho\psi_\sigma-\frac{i}{2}A_\sigma\gamma_5\psi_\rho+\frac{1}{2}\gamma_\sigma\eta\psi_\rho\right)~.
	\end{align}

The gauge condition (\ref{gravinogauge}) is understood from now as having been imposed in the fermionic sector of the supergravity action. 
As mentioned previously, it is in this gauge that the decoupling of the Goldstino from the supergravity action, upon an appropriate redefinition of the gravitino
field, occurs. \\

To investigate the possibility of dynamical mass generation for the gravitino we firstly introduce an auxiliary scalar field $\sigma$ to linearise the four-gravitino interaction term in equation \eqref{Rexpansion} via the equivalence
	\begin{align}	
		\frac{e}{2\kappa^2}R\left(e,\psi\right)=
		\frac{e}{2\kappa^2}\left(R\left(e\right)+\frac{11}{4}\kappa^4\left(\overline{\psi_\mu}\Gamma^{\mu\nu}\psi_\nu\right)^2+\dots\right) \sim
		\frac{e}{2\kappa^2}\, R\left(e\right) - e\,\sigma^2 - e\,\frac{\sqrt{11}}{2 \, \sqrt{2}} \, \kappa \,\sigma \,\left(\overline{\psi_\mu}\, \psi^\mu \right)+\dots ,
	\end{align}
which follows as a simple consequence of the subsequent Euler-Lagrange equation for $\sigma$. 
Taking into account the Goldstino-induced negative cosmological constant term arising from equation \eqref{va2b}, the relevant terms in equation \eqref{va3} are then
	\begin{align}\label{actionauxiliary}
		&\mathcal{L}_{eff}=
		-\frac{e}{2\kappa^2}\left(R(e) + 2\kappa^2 \, \left(f^2 - \sigma^2 \right) \right)
		-\frac{1}{2}\epsilon^{\mu\nu\lambda\rho}\overline{\psi_\mu}\gamma_5\gamma_\nu D_\lambda\psi_\rho
		+\frac{\sqrt{11}}{2\, \sqrt{2}}e\kappa\sigma \left(\overline{\psi_\mu}\, \psi^\mu\right)
		+\dots.
	\end{align}
Following the normalisation for the gravitino mass of \cite{Fradkin:1983mq}, 
	\begin{equation}\label{normmass}
	\frac{1}{2}\epsilon^{\mu\nu\lambda\rho}\overline{\psi_\mu}\gamma_5\gamma_\nu D_\lambda\psi_\rho
	+m \left(\overline{\psi_\mu}\Gamma^{\mu\nu}\psi_\nu\right)~,
	\end{equation}	
we note that, if $\sigma$ acquires a non-zero vacuum expectation value (vev) through the process of quantisation so that
$\sigma_c\equiv\langle \sigma \rangle \ne 0$,
then we have dynamically generated an effective mass of 
	\begin{equation}\label{ginomass}
		m = \sqrt{\frac{11}{2}}\, \kappa\, \sigma_c~, 
	\end{equation}
for the gravitino, thus breaking local supersymmetry. 
The other four-gravitino terms (which involve $\gamma_5$) can be neglected in this regard as they are not of canonical form for mass terms, as can the (subleading 
at one-loop order) gravitino-graviton interaction terms.\\

Furthermore, we may then define
	\begin{equation}\label{bare}
		\Lambda_0  \equiv \kappa^2 \, \left(- f^2  + \sigma_c^2\right)\;
		\Rightarrow\; \mathcal{L}_{eff}=-\frac{e}{2\kappa^2}\left(R\left(e\right)-2\Lambda_0\right)+\dots
	\end{equation}
so that we may identify $\Lambda_0$ with a tree-level cosmological constant \footnote{We may also consider contributions of $S$ and $P$ 
to give a tree-level cosmological constant which we can tune in such a way 
that such contributions are absorbed in $f^2$ in (\ref{bare}), so that $\Lambda_0$ is understood to contain such contributions.}.\\

Upon quantisation this bare cosmological constant is then dressed by quantum corrections to yield a renormalised cosmological constant $\Lambda$. 
In this sense, although de Sitter space-time may not be a solution of the bare equations of motion stemming from $\mathcal{N}=1$ 
supergravity lagrangian \eqref{lagrangian} in its broken symmetry phase, it may be the solution of a \emph{quantum effective action}, after the fluctuations of the metric and other (gravitino and ghost) fields are taken into account. \\

For the purposes of our analysis here, and following \cite{Fradkin:1983mq}, we assume that one may expand the one-loop effective 
action about a de Sitter background with a positive renormalised cosmological constant, $\Lambda > 0$, whose value will be determined
by minimisation of the one-loop effective action.  
It is however known that in supergravity an apparent (but not actual) gravitino mass term is required in the presence of a non-zero cosmological constant \cite{Townsend:1977qa} for the closure of the supersymmetry algebra, mandating that we consider the limit $\Lambda \to 0$ to ensure that any such gravitino mass generated is indeed physical. \\

It is important to stress at this point that since $\Lambda$ is equivalently the overall energy density at the level of the one-loop 
effective potential $V_{eff}$, we then furthermore require as a self-consistency condition that 
$\left\{\sigma_c\neq0,V_{eff}\left(\sigma_c\right)=0\right\}$: i.e. that the energy density (and thus $\Lambda$) 
is indeed zero at whatever non-trivial minima we find.
This is achieved in practice by tuning $f^2$, which is contained within the bare cosmological constant \eqref{bare}. \\
 
Our algorithm is therefore to find the one-loop effective action for this theory in $d=4$ (Euclidean) de Sitter space ($R_{\mu\nu} = \Lambda \, g_{\mu\nu} $),
before solving the corresponding effective equations in the limit that the effective cosmological constant $\Lambda$ vanishes, enabling a straightforward 
interpretation of any resultant gravitino mass. \\
 
Some technical but important remarks are in order at this stage. 
One may consider the excursion of our theory through de Sitter space purely in the spirit of 
Euclidean continuation, as appropriate for path integrals in a consistent quantisation of (super)gravity, allowing us to arrive back at a physical theory only
in the limit $\Lambda\to 0$.
We may also note a tension here in that whilst the gravitinos in \eqref{lagrangian} are Majorana, there are in fact no Majorana representations on$S^4$ 
(or equivalently, in $SO(5)$).
As however we are treating this continuation as a purely technical step, we will proceed with the understanding that we will arrive at a physical result 
only after completing these manipulations.

\section{One-loop partition functions \label{sec:quadr}}
To compute the one-loop effective action for a given theory there are a number of operations that we must firstly take account of. 
Firstly, we must expand about a classical background to compute fluctuations of the action to quadratic order, via decompositions of 
the type $\tilde g_{\mu\nu}\to g_{\mu\nu}+h_{\mu\nu}$, where for our purposes $g_{\mu\nu}$ is the standard Euclidean $dS_4$ metric. 
Working to one-loop order in this instance has the advantage of decoupling the gravitino and graviton sectors, as all gravitino terms 
in \eqref{lagrangian} are already quadratic, and Lorentz invariance forbids any fermionic background terms.\\

Since we are interested in the one-loop effective potential for the auxiliary field $\sigma$, it is sufficient to identify the latter with its
vev $\sigma\to\sigma_c$. Indeed, the auxiliary field does not propagate at tree level, and its kinetic term (obtained from the integration over other degrees
of freedom, for a non-uniform configuration $\sigma$) would therefore be purely one-loop. Hence its influence on the effective potential, which is already 
of one-loop order, would be at least of two-loop order. We therefore replace from now on $\sigma$ by $\sigma_c$. 
Note that the effective potential obtained in this way would be exact if one integrated over $N\to\infty$ fermions (which is excluded here because of 
the matching of bosonic and fermionic degrees of freedom).\\

We must also decompose all fields present (including ghosts) into those that are `natural' to our background geometry; more precisely, 
to those  corresponding to irreducible representations of the underlying isometry group.
It is the spectra of these operators that can be reliably found through knowledge of the underlying representation theory, which will then 
allow us to compute the effective action.

\subsection{Bosonic sector}
Starting with the gravitational lagrangian in \eqref{bare}, we firstly vary to quadratic order in $h_{\mu\nu}$, yielding (in the metric formalism) \cite{Fradkin:1983mq}
	\begin{align}\label{GRaction}
		\frac{1}{4\kappa^2}\int d^4 x\sqrt{g}\left[\frac{1}{2} \overline h_{\mu\nu}\left(-\nabla^2
		+\frac{8}{3}\Lambda-2\Lambda_0\right) \overline h^{\mu\nu}
		-\frac{1}{8}h\left(-\nabla^2-2\Lambda_0\right)h
		-\left(\nabla^\mu  \overline h_{\mu\nu}-\frac{1}{4}\nabla_\nu h\right)^2\right]~,
	\end{align}
where we have followed the convention that $D$ and $\nabla$ are the spin and Christoffel connections respectively, 
and have also made the standard decomposition $h_{\mu\nu}=\overline h_{\mu\nu}+g_{\mu\nu} h/4$. \\

It is important to also note that given the presence of fermions we must work in the vierbein formalism, which leads to extra terms (which vanish on-shell) in the quadratic action relative to the metric formalism \cite{Percacci:2013ii}. 
This can be understood as arising from the first variation of the gravitational action, which takes the form
	\begin{align}
		\left(G^{\mu\nu}+g^{\mu\nu}\Lambda_0\right)\delta g_{\mu\nu}
		=\left(G^{\mu\nu}+g^{\mu\nu}\Lambda_0\right)\delta e_{(\mu}{}^a\eta_{|ab|}e_{\nu)}{}^b~.
	\end{align}
We can then see that the second variation in the vierbein formalism will coincide with the metric formalism, along with the additional term
	\begin{align}
		&\frac{1}{4\kappa^2}\int d^4 x\sqrt{g}\left(G^{\mu\nu}+g^{\mu\nu}\Lambda_0\right)\delta e_{(\mu}{}^a\eta_{|ab|}\delta e_{\nu)}{}^b
		=\frac{1}{16\kappa^2}\int d^4 x\sqrt{g}\;g^{\mu\nu}\left(G+4\Lambda_0\right)\delta e_{(\mu}{}^a\delta e_{\nu) a}\nonumber\\
		&=\frac{1}{16\kappa^2}\int d^4 x\sqrt{g}\;\left(\Lambda_0-\Lambda\right)h_{\mu\nu}^2
		=\frac{1}{4\kappa^2}\int d^4 x\sqrt{g}\;\left(\frac{\Lambda_0-\Lambda}{2}\right)\left(\frac{1}{2}\overline{h}_{\mu\nu}^2+\frac{1}{8}h^2\right)~,
	\end{align}
where we have used that $h_{\mu\nu}=2e_{(\mu}{}^a\delta e_{\nu) a}$ and that $g^{\mu\nu}=e^{(\mu}{}_a\eta^{|ab|} e^{\nu)}{}_b$.	\\

We therefore re-express \eqref{GRaction} as
	\begin{align}
		&\frac{1}{4\kappa^2}\int d^4 x\sqrt{g}\left[\frac{1}{2} \overline h_{\mu\nu}\left(-\nabla^2+X_1\right) \overline h^{\mu\nu}
		-\frac{1}{8}h\left(-\nabla^2-X_2\right)h
		-\left(\nabla^\mu  \overline h_{\mu\nu}-\frac{1}{4}\nabla_\nu h\right)^2\right]~,\\	
		&X_1=\frac{8}{3}\Lambda-2\Lambda_0+\frac{\Lambda_0-\Lambda}{2}=\frac{13}{6}\Lambda-\frac{3}{2}\Lambda_0, \quad
		X_2=2\Lambda_0+\frac{\Lambda_0-\Lambda}{2}=\frac{5}{2}\Lambda_0-\frac{1}{2}\Lambda~.
	\end{align}

To further decompose into the `irreducible action' we will make use of the standard `transverse traceless' decomposition
	\begin{align}
		&\label{eq2} V_\mu=V_\mu^\perp+\nabla_\mu\phi~, \quad 
		\nabla^\mu V_\mu^\perp=0~,\quad
		\mathcal{D}V=\mathcal{D}V^\perp\mathcal{D}\phi\sqrt{\det\Delta_0\left(0\right)}\\
		&\label{decomp2}
		\overline h_{\mu\nu}=\overline h_{\mu\nu}^\perp+\nabla_\mu\xi_\nu^\perp+\nabla_\nu\xi_\mu^\perp
		+\nabla^2_{\mu\nu}\chi
		-\frac{1}{4}g_{\mu\nu}\nabla^2\chi~,\quad
	 	g^{\mu\nu}\overline h_{\mu\nu}=0~,\\
		&\nabla^\nu\overline h_{\mu\nu}^\perp=0~,\quad
		\mathcal{D}\overline{h}_{\mu\nu}
		=\mathcal{D}\overline{h}^\perp\mathcal{D}\xi^\perp_\mu\mathcal{D}\chi\sqrt{\det\Delta_1\left(-\Lambda\right)\otimes\Delta_0\left(-\frac{4}{3}\Lambda\right)\otimes\Delta_0\left(0\right)}~,
	\end{align}
where $V_\mu$ is some vector field and we have also defined a class of bosonic operators for constant $X$
	\begin{align}\label{decompositions1}
		\Delta_0(X)\;\phi&=\left(-\nabla^2+X\right)\phi~,\\
		\Delta_1^{\mu\nu}(X)\;\xi_\nu^\perp&=\left(-\nabla^{2\mu\nu}+g^{\mu\nu}X\right)\xi_\nu^\perp~,\\
		\Delta_{2\alpha\beta}^{\mu\nu}(X)\;\overline h_{\mu\nu}^\perp&=\left(-\nabla^{2\mu\nu}_{\alpha\beta}
		+\delta_\alpha^\mu\delta_\beta^\nu X\right)\overline h_{\mu\nu}^\perp~.
	\end{align}
It is these operators whose spectra we shall ultimately compute. 
There is however an important caveat that we must bear in mind in that are extra zero-modes present in these decomposed operators, which must be correctly accounted for \cite{Fradkin:1983mq}.
We will take account of these later, where we will note that in the limit $\Lambda\to0$ their contributions are of subleading order.\\

To streamline the process of computation we will make use of the following ($S^4)$ identities 
	\begin{align}					  	
		&V^\mu\left(\left(-\nabla^2+X\right)\delta_{\mu\nu}+c\nabla_{\mu\nu}\right)V^\nu
		=V_\mu^\perp\Delta_1\left(X\right)V^{\mu\perp}
		+\left(1-c\right)\phi\Delta_0\left(0\right)\Delta_0\left(\frac{X-\Lambda}{{1-c}}\right)\phi~,
	\end{align}
\begin{align}
	\overline{h}^{\mu\alpha}&\left(\left(-\nabla^2+X\right)\delta_{\mu\nu}+c\nabla_{\mu\nu}\right)\overline{h}^{\nu}{}_\alpha=\\\nonumber
	&\qquad
	\begin{pmatrix}
		\overline{h}^{\perp\mu\alpha}&&
		\xi^{\mu\perp}&&
		\chi
	\end{pmatrix}\cdot
	\begin{pmatrix}
		\Delta_2\left(X\right)\\
		\left(2-c\right)\Delta_1\left(-\Lambda \right)\Delta_1\left(\frac{3 c \Lambda -10 \Lambda +6 X}{6-3 c}\right)\\
		\frac{3}{16}\left(4-3c\right)\Delta_0\left(0\right)\Delta_0\left(-\frac{4 \Lambda }{3}\right)
		\Delta_0\left(\frac{4 ((3 c-8) \Lambda +3 X)}{12-9 c}\right)
	\end{pmatrix}_{\text{Diag.}}\cdot
	\begin{pmatrix}
		\overline{h}^{\perp}_{\mu\alpha}\\
		\xi_\mu^\perp\\
		\chi
	\end{pmatrix}~,
\end{align}
for some constants $X$ and $c$.

\subsubsection{Gauge fixing}
We have two symmetries present in this sector; local Lorentz and infinitesimal coordinate transformations.
To fix the former we simply follow the convention of setting the antisymmetric part of the vierbein to zero \cite{VanNieuwenhuizen:1981ae}, leading to non-propagating ghost fields which can then be disregarded here.
To fix the coordinate gauge transformations we add a standard two parameter covariant gauge fixing term
	\begin{align}\label{bosonic gauge fixing}
		S_B^{(GF)}=-\frac{1}{4\kappa^2}\frac{1}{\alpha}\int d^4 x\sqrt{g}\left(\nabla^\mu  h_{\mu\nu}-\frac{1+\beta}{4}\nabla_\nu h\right)^2~,
	\end{align}
which necessitates the ghost action
	\begin{align}
		S_B^{(GH)}=\frac{1}{4\kappa^2}\frac{1}{\alpha}\int d^4 x\sqrt{g}\;
		\overline{C}^\mu\left(\left(-\nabla^2-\Lambda\right)\delta_{\mu\nu}+\frac{\beta-1}{2}\nabla_{\mu\nu}\right)C^\nu~,
	\end{align}
for some anticommuting complex vector field $C$.\\

This may be easily integrated after applying the first identity given previously to arrive at the ghost partition function
	\begin{align}
		\mathcal{Z}^{(GH)}_B
		=\det\Delta_1\left(-\Lambda\right)\otimes\Delta_0\left(\frac{4 \Lambda }{\beta -3}\right)~,
	\end{align}
where we have absorbed any prefactors into the normalisation of the functional measure.

\subsubsection{Physical gauge}\label{sec:physical gauge}

There is however a secondary line of approach by which we may instead address the issue of gauge fixing. 
This is via an appeal to so-called `physical' gauges; gauges which represent an alternative path to quantisation than the conventional Faddeev-Popov method, inasmuch as they consist of isolating the gauge degrees of freedom present and essentially disregarding them as a `physical' gauge choice. \\

In practice this is achieved in the following manner.
It is firstly well known that we must gauge fix in order to render path integrals well defined.
Without dividing out by the volume of the gauge group, we naturally overcount field configurations which are physically equivalent and related by gauge transformations.
Conventionally this is remedied via a gauge fixing condition specifying a phase-space hypersurface which intersects each orbit of the gauge group, along with the inclusion of a Faddeev Popov ghost determinant in the path integral measure to locally cancel the phase-space curvature of non-Abelian gauge symmetries.
This is however not the only path we may proceed by.\\

Re-examining our gauge fixing condition \eqref{bosonic gauge fixing} we may consider the illuminative case $\alpha\to0$, whereupon we may strongly impose (i.e. use at the level of the action, rather than solely imposing via Lagrange multiplication) the condition
	\begin{align}
		\nabla_\mu h^\mu{}_\nu-\frac{\beta+1}{4}\nabla_\nu h=0~.
	\end{align}
Substituting the decomposition \eqref{decomp2} and setting $\beta=0$, this condition then becomes $\nabla_\mu\nabla^\mu\xi_\nu+\nabla_\mu\nabla_\nu\xi^\mu=0$ for $\xi_\mu=\xi_\mu^{T}+\nabla_\mu\chi$. 
We may then straightforwardly note that $\xi_\mu$ must therefore be Killing, and as such, it is $\xi_\mu$ that parametrises the underlying diffeomorphism symmetry present.
Our `physical' gauge condition is then to strongly impose that $\xi_\mu=0$ and therefore disregard the components of the graviton corresponding to general coordinate transformations; i.e. the fields $\xi_\mu^T$ and $\chi$ in our notation.
The functional integral over these gauge degrees of freedom then yields an infinite constant prefactor of the volume of the diffeomorphism group, which is unimportant for our purposes.\\

We should note at this point that there are some additional complications to this physical gauge procedure regarding the correct counting of the zero modes of $\xi_\mu$ (which should not necessarily be disregarded even if $\xi_\mu=0$), however as in the previous instance of the extra zero modes arising from our decompositions, these contributions are subleading as $\Lambda\to0$. \\

Given the inevitable gauge dependence of effective potentials such as these; a consequence of our artificial truncation to one-loop order, it is arguable that physical gauge techniques are more `natural' in this context, and certainly can offer significant computational simplifications.
It is perhaps unsurprising then that they have been utilised effectively in a number of situations similar to these \cite{Percacci:2013ii}\cite{Gaberdiel:2010xv}\cite{Zhang:2012kya}.\\

Naturally of course we may apply these same considerations to the gauge fixing of the fermionic sector, however for our purposes it will suffice to utilise this procedure only in the bosonic case.
As we will see, in this context it is ultimately the behaviour of the bosonic sector that dictates the stability of the effective potential.

\subsubsection{Gravitational partition function}
With the two extra contributions outlined above, along with our gauge fixing term, the quadratic gravitational action now becomes
	\begin{align}\label{gauge fixed action}
		S_B^{(2)}=\frac{1}{4\kappa^2}\int d^4 x\sqrt{g}\bigg[&\frac{1}{2} \overline h_{\mu\nu}\left(\left(-\nabla^2
		+X_1\right)\delta^{\mu}{}_\alpha+2\left(1+\frac{1}{\alpha}\right)\nabla^{\mu}{}_\alpha\right) \overline h^{\nu\alpha}\\\nonumber
		&-\frac{\left(3\alpha+\beta^2\right)}{16\alpha}h\left(-\nabla^2-\frac{2X_2\alpha}{\left(3\alpha+\beta^2\right)}\right)h
		-\frac{\alpha+\beta}{2\alpha}\overline{h}_{\mu\nu}\nabla^{\mu\nu}h\bigg]~,
	\end{align}

Performing the functional integral would be straightforward, were it not for the final term in \eqref{gauge fixed action}. 
However, we may firstly note from \eqref{decompositions1} that $\overline{h}^\perp_{\mu\nu}$ is conserved and so cannot mix with $h$.
Furthermore, we may also leverage the result that for Einstein backgrounds there is no $\left(\xi^\perp,h\right)$ mixing \cite{Lauscher:2001ya}.
To eliminate the final $\left(\chi,h\right)$ mixing we could then also use a so-called `diagonal gauge' \cite{Percacci:2013ii}, however this would be incompatible with the Landau-DeWitt gauge choice, which we know to correspond to the unique gauge-invariant one-loop effective action for pure Einstein gravity \cite{Fradkin:1983nw}.\\

Instead, we will proceed with our general-gauge calculation. 
Schematically, the scalar part of the action is then of the form
	\begin{align}
		\frac{1}{4\kappa^2}\int d^4 x\sqrt{g}\left(
		\begin{pmatrix}
			h && \chi
		\end{pmatrix}
		\cdot
		\begin{pmatrix}
			A1 && B\\
			B && A2
		\end{pmatrix}
		\cdot
		\begin{pmatrix}
			h \\ 
		\chi
		\end{pmatrix}\right)
	\end{align}
with matrix elements	
	\begin{align}
		A_1&=-\frac{1}{16\alpha}\left(-\left(3\alpha+\beta^2\right)\nabla^2-2X_2\alpha\right),\\
		A_2&= -\frac{3 (\alpha +3)}{16 \alpha}\Delta_0\left(0\right)\Delta_0\left(-\frac{4}{3}\Lambda\right)\Delta_0\left(\frac{4 (\alpha -3) \Lambda -6 \alpha  X_1}{3 (\alpha +3)}\right),\\
		B&=-\frac{3\left(\alpha+\beta\right)}{16\alpha}\Delta_0\left(0\right)\Delta_0\left(-\frac{4}{3}\Lambda\right)~,
	\end{align}
so that we may integrate this along with the other fields present to find the bosonic partition function
	\begin{align}
		\mathcal{Z}^{(B)}&=\mathcal{Z}^{(GH)}_B\otimes
		\left(\frac{\det\Delta_0\left(-\frac{4}{3}\Lambda\right)\otimes\Delta_0\left(0\right)}
		{\det\Delta_2\left[X_1\right]\otimes\Delta_1\left(\alpha\left(\frac{2 }{3}\Lambda-X_1\right)-\Lambda\right)\otimes
		\left(A_1A_2-B^2\right)}\right)^{1/2}\nonumber\\
		&=\det\Delta_1\left(-\Lambda\right)
		\otimes\Delta_0\left(\frac{4 \Lambda }{\beta -3}\right)
		\otimes\left(\Delta_2\left(X_1\right)
		\otimes\Delta_1\left(\alpha\left(\frac{2}{3}\Lambda-X_1\right)-\Lambda\right)
		\otimes\Delta_0\left(\frac{A_3\pm \sqrt{A4}}{6\left(\beta-3\right)^2}\right)\right)^{-1/2}~,
	\end{align}
	\begin{align}
		A_3&= 4\Lambda  \left(6 \alpha +\beta^2+6 \beta -9\right)
		-6 X_1 \left(3 \alpha +\beta^2\right)
		+6(\alpha +3) X_2~,\\
		A_4&=4 \left(2 \Lambda(6\alpha+\beta(\beta +6)-9)
		-3X_1 \left(3\alpha+\beta^2\right)
		+3(\alpha +3) X_2\right)^2
		+48 (\beta -3)^2 X_2 (3 \alpha X_1-2 (\alpha -3)\Lambda )~,
	\end{align}
where we have again disregarded an irrelevant multiplicative prefactor.\\

As a quick check we may verify that we can reproduce known results from the literature.
If we consider the replacements $\left\{X_1\to\frac{8}{3}\Lambda-2\Lambda_0,X_2\to2\Lambda_0,\beta\to1,\alpha\to0\right\}$, corresponding to Einstein gravity in Landau-DeWitt gauge, we find 
	\begin{align}\label{grlimit}
		\mathcal{Z}^{(B)}&\to\left(\frac{\det\Delta_1\left(-\Lambda\right)\otimes\Delta_0\left(-2\Lambda\right)}
		{\det\Delta_2\left(\frac{8}{3}\Lambda-2\Lambda_0\right)
		\otimes\Delta_0\left(-2\Lambda_0\right)}\right)^{1/2}~,
	\end{align}
arriving precisely at the partition function given in \cite{Fradkin:1983mq}.
Equivalent results in other gauges follow similarly. 

\subsection{Fermionic sector}
On the fermionic side we follow largely the same approach as utilised in the previous section, with the exception that rather than starting from the Euclidean ($S^4$) action, we will utilise \eqref{lagrangian} and perform the continuation at an opportune moment.
We have the action
	\begin{align}\label{fermion action}
		S_F^{\left(2\right)}
		=\int d^4 x\sqrt{-g}\left(-\frac{1}{2}\epsilon^{\mu\nu\lambda\rho}\overline{\psi_\mu}\gamma_5\gamma_\nu D_\lambda\psi_\rho
		-\frac{\sqrt{11}}{\sqrt{2}}\kappa\sigma_c \left(\overline{\psi_\mu}\Gamma^{\mu\nu}\psi_\nu\right)
		\right)~,
	\end{align}
which, given the absence of fermionic background terms, is already quadratic in quantum fields.
From the standard decompositions	
	\begin{align}\label{eq1} 
		\psi_\mu&=\varphi_\mu+\frac{1}{4}\gamma_\mu\psi~, \quad 
		\gamma^\mu\varphi_\mu=0~,\quad
		\varphi_\mu=\varphi_\mu^\perp+\left(D_\mu-\frac{1}{4}\gamma_\mu\slashed{D}\right)\zeta~, 
		\quad D^\mu\varphi_\mu^\perp=0~,\quad
		\mathcal{D}\psi_\mu=\frac{\mathcal{D}\varphi^\perp_\mu\mathcal{D}\psi\mathcal{D}\zeta}{\sqrt{\det\Delta_{1/2}\left(-\frac{4}{3}\Lambda\right)}}~,
	\end{align}
where $\psi=0$ in our gauge choice \eqref{gravinogauge}.
We then define a class of fermionic operators for constant $X$
	\begin{align}\label{decompositions}
		\Delta_{1/2}(X)\;\psi&
		=\left(-D^2+\Lambda+X\right)\psi~,\quad
		\Delta_{3/2}^{\mu\nu}(X)\;\varphi_\mu^\perp
		=\left(-D^{2\mu\nu}+\frac{4}{3}\Lambda g^{\mu\nu}+g^{\mu\nu}X\right)\varphi_\mu^\perp~,
	\end{align}
where explicit $\Lambda$ terms are largely for future convenience and coherence with the literature.\\

We Euclideanise via the transformations
	\begin{align}
		\left\{\gamma^0\to i\gamma^0_E~, \quad
		\gamma^j\to \gamma^j_E~, \quad
		e^0\to e^0_E~, \quad 
		e^j\to ie^j_E\right\}\Rightarrow
		\slashed{D}\to i\slashed{D}_E~,
	\end{align}
which, since the Dirac operator `squares' to give the Laplacian
	\begin{align}
		\slashed{D}_E^2=-D^2+\frac{R}{4}~,
	\end{align}
provides the useful transformation 
	\begin{align}\label{dirac square}
		-\slashed{D}^2\to-D^2+\frac{R}{4}~,
	\end{align}
which we may then apply to simultaneously remove $\slashed{D}$ operators and Euclideanise the theory.\\

As before, we may streamline computations via the $(S^4)$ identity given in \cite{Fradkin:1983mq} (using \eqref{dirac square} as appropriate) 
	\begin{align}
		&\frac{1}{2}\epsilon^{\mu\nu\lambda\rho}\overline{\psi_\mu}\gamma_5\gamma_\nu D_\lambda\psi_\rho
		+m \left(\overline{\psi_\mu}\Gamma^{\mu\nu}\psi_\nu\right)\\
		&=\frac{1}{2}\overline{\varphi}^\perp_\mu\left(\slashed{D}-m\right)\varphi^{\perp\mu}
		+\frac{3}{16}
		\begin{pmatrix}
			\overline\zeta && \overline\psi
		\end{pmatrix}\cdot
		\begin{pmatrix}
			\left(\slashed{D}+2m\right)\Delta_{1/2}\left(-\frac{4}{3}\Lambda\right) && -\Delta_{1/2}\left(-\frac{4}{3}\Lambda\right)\nonumber \\
			-\Delta_{1/2}\left(-\frac{4}{3}\Lambda\right) && -\left(\slashed{D}-2m\right)
		\end{pmatrix}\cdot 
		\begin{pmatrix}
			\zeta \\ 
			\psi
		\end{pmatrix}~,
	\end{align}
where $m$ is a generic mass term.

\subsubsection{Gauge fixing}

Although we have already imposed the gauge condition via \eqref{gravinogauge}, to implement it consistently (and with a view to preserving local supersymmetry) we may firstly consider a more general gauge fixing strategy (which will in fact supersede the condition \eqref{gravinogauge}), before specialising to the specific instance of $\gamma\cdot\psi=0$.\\

In generality, to fix the local supersymmetry present in the gravitino sector we must supplement \eqref{fermion action} with some gauge-fixing term, which, to preserve on-shell supersymmetry, we may derive via considerations of the variation of \eqref{bosonic gauge fixing} with respect to the transformation $\delta h_{\mu\nu}=\kappa\bar{\epsilon}\gamma_{(\nu}\psi_{\mu)}$ (where $\epsilon$ is assumed to be Killing, thus obeying the $S^4$ relation $D_\mu\epsilon=\frac{1}{2}\sqrt{-\frac{R}{12}}\gamma_\mu\epsilon$).
Note that for an dS space in our conventions $R>0$, and thus the latter relation can contribute imaginary terms, these vanish in the limit $\Lambda\to0$ that we consider herein.

As encountered in other circumstances, strict proportionality between the fermionic and bosonic gauge fixing terms is difficult due to the presence of $\varphi^\perp_\mu$ terms in the variation of \eqref{bosonic gauge fixing} \cite{Percacci:2013ii,Kojima:1993hb}.
As a compromise however, we may find a proportionality in the following manner.
Taking the variation and the subsequent $\gamma$-trace (and noting that we will apply the eventual constraint $\gamma\cdot\psi=\psi=0$), we find
	\begin{align}\label{fermion gauge 1}
		\gamma^\mu\left(\nabla_{\nu}\delta h^\nu{}_\mu+\frac{\beta+1}{4}\nabla_\mu \delta h\right)
		= \frac{3}{2}\kappa\bar{\epsilon}\left(D^2+\frac{R}{12}\right)\zeta~,
	\end{align}
which suggests the following gauge fixing term	 
	\begin{align}
		S_F^{(GF)}=\frac{1}{2}\int d^4 x\sqrt{-g}\;\overline{F} \left(\slashed{D}+\sqrt{-\frac{R}{3}}\right)F~, \quad 
		F= \left(\slashed{D}-\sqrt{-\frac{R}{3}}\right)\zeta~,
	\end{align}
since
	\begin{align}\label{fermion gauge 2}
		\left(\slashed{D}+\sqrt{-\frac{R}{3}}\right)F
		= \left(D^2+\frac{R}{12}\right)\zeta~,
	\end{align}
and on-shell proportionality between \eqref{fermion gauge 1} and \eqref{fermion gauge 2} is ensured. \\

To now find the corresponding ghost action we vary $F$ about the classical background (where $S=P=A_\mu=0$) by decomposing $\delta\psi_\mu$ in \eqref{susy transformations} to give
	\begin{align}
		\left(D^2-\frac{R}{12}\right)\left(\delta\zeta-\frac{\epsilon}{\kappa}\right)=0~,
	\end{align} 
which yields the ghost action
	\begin{align}
			S_F^{(GH)}=\frac{1}{\kappa}\int d^4 x\sqrt{-g}\;\overline\eta\left(\slashed{D}-\sqrt{-\frac{R}{3}}\right)\eta~,
	\end{align}
for some commuting complex spin $1/2$ field $\eta$.\\

To ensure on-shell gauge independence we must also take account of so-called third (or Nielsen-Kallosh) ghosts arising from the non-trivial $\left(\slashed{D}+\sqrt{-\frac{R}{3}}\right)$ operator in our gauge fixing condition.
Exponentiating, we find
	\begin{align}
		S_F^{(NK)}=\int d^4 x\sqrt{-g}\;\left(\overline\omega \left(\slashed{D}+\sqrt{-\frac{R}{3}}\right)\omega
		+\overline\rho \left(\slashed{D}+\sqrt{-\frac{R}{3}}\right)\rho\right)~,
	\end{align}
for some anticommuting Majorana and commuting Dirac spinor fields $\omega$ and $\rho$, respectively.	\\

Integrating gives the fermionic ghost partition function
	\begin{align}
		\mathcal{Z}^{(GH)}_F=\left(\det\left(\slashed{D}-\sqrt{-\frac{R}{3}}\right)\right)^{-1}\left(\det\left(\slashed{D}+\sqrt{-\frac{R}{3}}\right)\right)^{-1/2}
		=\left(\det\Delta_{1/2}\left(-\frac{R}{3}\right)\right)^{-3/4}~,
	\end{align}
where we have leveraged \eqref{dirac square} to equate
	\begin{align}
		\det\left(\slashed{D}\pm m\right)
		=\left(\det\Delta_{1/2}\left(m^2\right)\right)^{1/2}~,
	\end{align}
which, as in the bosonic case, is true modulo the additional zero modes incurred by the decomposition we have used.

\subsubsection{Gravitino partition function}

Combining these elements and noting that the gauge condition \eqref{gravinogauge} implies the vanishing of the spin 1/2 field $\psi$, we find the quadratic gravitino terms
	\begin{align}
		-\frac{1}{2}\overline{\varphi}^\perp_\mu\left(\slashed{D}-\frac{\sqrt{11}}{\sqrt{2}}\kappa\sigma_c\right)\varphi^{\perp\mu}
		-\frac{11}{16}\overline{\zeta}\left(\slashed{D}+\frac{3\sqrt{2}}{\sqrt{11}}\kappa\sigma_c-\frac{8}{11}\sqrt{-\frac{4}{3}\Lambda}\right)\Delta_{1/2}\left(-\frac{4}{3}\Lambda\right)\zeta~,
	\end{align}
which we can integrate as before to give the total fermionic partition function (including Jacobian factors)
	\begin{align}
		\mathcal{Z}^{(F)}
   		&=\left(\frac{\det\Delta_{3/2}\left(\frac{11}{2}\kappa^2\sigma_c^2\right)\otimes\Delta_{1/2}
   		\left(\left(\frac{3\sqrt{2}}{\sqrt{11}}\kappa\sigma_c-\frac{8}{11}\sqrt{-\frac{4}{3}\Lambda}\right)^2\right)}
		{\left(\det\Delta_{1/2}\left(-\frac{4}{3}\Lambda\right)\right)^3}\right)^{1/4}~,
	\end{align}
and we have again leveraged \eqref{dirac square} to equate
	\begin{align}
		\left(\det\left(\slashed{D}\pm m\right)_{\varphi^\perp}\right)
		=\left(\det\Delta_{3/2}\left(m^2\right)\right)^{1/2}~,
	\end{align}
which, as in the bosonic case, is true modulo the additional zero modes incurred by the decomposition we have used.

\section{One-loop effective potential and symmetry breaking patterns \label{sec:potential}}

Having derived the relevant partition functions, we are now in a position to compute the one-loop effective action via the relation
	\begin{align}
		\Gamma=-\ln \left(\mathcal{Z}^{(B)}\otimes \mathcal{Z}^{(F)}\right)
		=\frac{1}{2}\ln\det\Delta_{2}\left(X_1\right)+\dots,
	\end{align}
in conjunction with the functional determinant techniques detailed in the appendix, i.e. 
	\begin{align}\label{mudef}
		\ln\det\left(\frac{\Delta_s}{\mu^2}\right)=
		-\frac{1}{2}B_0 L^4-\frac{1}{2}B_2 L^2
		-B_4\left(\ln\left(\frac{L^2}{\mu^2}\right)
		-\gamma\right)
		+B_4\ln\left(\frac{\Lambda}{3\mu^2}\right)
		-\zeta_s'\left(0\right)~,
	\end{align}
where we have a cut-off $\epsilon=\left(\mu^2/L^2\right)\to0$, $\gamma$ is the Euler-Mascheroni constant, and the operator $\Delta_s$ is `non-decomposed'.	
The mass dimension of $\mu$ and $L$ is one, and it is important to note at this point that as $\epsilon$ arises from equation \eqref{Mellin} in the Appendix, which can be thought of as a proper time integral, $\epsilon\to0$ is a short time (and thus high energy) cutoff.
Flowing from the UV to IR therefore corresponds to the direction of increasing $\mu^2$.\\

Given that we are investigating a simple model of supergravity which we anticipate the embedding thereof in the context of a more UV-complete theory, we set aside the renormalisability of the action for now (encoded in the divergent terms above (as an aside, $B_0$ is always zero for supersymmetric theories)). 
Instead, we will focus on the finite parts of our `decomposed' effective potential $V_\text{eff}$ and the resultant effective equations. 
We may represent the (finite parts of the) `decomposed' effective potential as
	\begin{align}\label{one-loop action}
		\Gamma=&S_{c}
		+\left(B_4-N\right)\ln\left(\frac{\Lambda}{3\mu^2}\right)
		-B_4'~, \quad 
		N=14-\frac{1}{2}\times 8=10~,\quad
		V_\text{eff}=-\frac{\Lambda^2}{24\pi^2}\Gamma~,
	\end{align}
where $24\pi^2/\Lambda^2$ is the usual spacetime volume for an $S^4$ of radius $\sqrt{3/\Lambda}$, and 
	\begin{align}
		S_{c}=-\frac{1}{2\kappa^2}\int d^4x\;\sqrt{g}\left(R-2\Lambda_0\right)
		=-\frac{12\pi^2}{\Lambda^2\kappa^2}\left(R-2\Lambda_0\right)~,	
	\end{align}
	\begin{align}\nonumber\label{B_4}
		B_4=&\frac{1}{2}\zeta_2\left(0,X_1\right)
		-\frac{1}{4}\zeta_{3/2}\left(0,\frac{11}{2}\kappa^2\sigma_{c}^2\right)
		-\zeta_1\left(0,-\Lambda\right)
		+\frac{1}{2}\zeta_1\left(0,\alpha\left(\frac{2}{3}\Lambda-X_1\right)-\Lambda\right)\\
		&-\frac{1}{4}\zeta_{1/2}\left(0,\left(\frac{3\sqrt{2}}{\sqrt{11}}\kappa\sigma_c-\frac{8}{11}\sqrt{-\frac{4}{3}\Lambda}\right)^2\right)
		+\frac{3}{4}\zeta_{1/2}\left(0,-\frac{4}{3}\Lambda\right)
		-\zeta_{0}\left(0,\frac{4\Lambda}{\beta-3}\right)
		+\frac{1}{2}\zeta_{0}\left(0,\frac{A_3\pm \sqrt{A4}}{6\left(\beta-3\right)^2}\right)~,
	\end{align}
	\begin{align}\label{B_4'}
		B_4'=\frac{1}{2}\zeta_2'\left(0,X_1\right)+\dots,\quad
		\zeta_s\left(0,X\right)\bigg|_{\Lambda\to0}\sim\frac{6s+3}{4}\frac{X^2}{\Lambda^2}~,\quad
		\zeta'_s\left(0,X\right)\bigg|_{\Lambda\to0}\sim\frac{6s+3}{4}\frac{X^2}{\Lambda^2} \left(\frac{3}{2}-\ln\left(\frac{3X}{\Lambda}\right)\right)~,
	\end{align}
where $N$ is the number of extra zero modes incurred by our decompositions (as alluded to previously, and first elucidated in \cite{Fradkin:1983mq}).\\

We may note at this point the importance of the asymptotic forms for $\zeta$ and $\zeta'$ presented above (and derived in the appendix); it is these relations that allow us to express the effective potential in terms of elementary functions, rather than the awkward integrals of digamma functions from which they derive in this instance, and thus fully investigate the behaviour of $V_\text{eff}$.\\

We should also note that one may be alarmed by the presence of an $\mathcal{O}\left(\ln\left(\Lambda\right)\Lambda^{-2}\right)$ term in \eqref{B_4'}, which would naturally dominate over any classical contributions as $\Lambda\to0$, suggesting a failure of our one-loop approach.
We may however note that there is a precise cancellation of any such terms at the level of the effective potential.
More specifically, for every $\zeta'_s$ in \eqref{one-loop action} there is a corresponding $\zeta_s$ carrying the same coefficient and opposite sign.
As $\zeta_s$ also gains a factor of $\ln\left(\Lambda\right)$ in \eqref{one-loop action}, this then cancels exactly with the corresponding term arising from \eqref{B_4'}.

\subsection{Imaginary terms}

One may also naturally be alarmed by the presence of $\ln\left(X\right)$ terms in the above, as $X$ has the potential to become negative.
This would then yield imaginary terms in the effective potential, which, if not artefacts of our one-loop formalism, would indicate an instability and thus the impossibility of dynamical gravitino condensation in this context.
We may address this issue in two distinct ways.\\

Given that the parameter we have available to tune is $f$, we will consider the case where $\Lambda_0<0$: i.e. whereupon after varying $\sigma_c$ we self-consistently find non-trivial minima $\sigma_c^2\leq f^2$.
This is natural in the present context as we are considering a renormalised cosmological constant $\Lambda=\Lambda_0+\mathcal{O}\left(\hbar^2\right)$, and it is known generically that quantisation of metric fluctuations about $dS_4$ leads to (positive) Planckian values for $\Lambda$ \cite{Fradkin:1983mq}.
To thus arrive at the case $\Lambda\sim0$ we must tune $\Lambda_0$ (\ref{bare}) to cancel out the positive energy density incurred via quantisation, mandating that $\Lambda_0<0$.\\

Given that for $\Lambda\to0$ $X_1\to-3\Lambda_0/2$, it is firstly straightforward to note upon inspection of \eqref{B_4} that in the limit $\Lambda\to0$ the only $\zeta'$ functions from \eqref{B_4'} which may be problematic in this sense are 
	\begin{align}\label{zetas2}
		\left\{\zeta'_1\left(0,\frac{3}{2}\alpha\Lambda_0\right),\quad
		\zeta'_{0}\left(0,\frac{A_3\pm \sqrt{A4}}{6\left(\beta-3\right)^2}\right)\right\}~,
	\end{align}
corresponding to the fields $\xi_\mu^T$, $\chi$ and $h$.\\
	
Working for convenience in the gauge $\alpha\to0$, these become
	\begin{align}\label{zetas}
		\left\{\zeta'_1\left(0,0\right)~,\quad
		\zeta'_{0}\left(0,\frac{3 \left(\beta ^2+5\right)\Lambda_0}{(\beta -3)^2}\right)~,\quad
		\zeta'_{0}\left(0,0\right)\right\}~,
	\end{align}
and it is simple to note that via \eqref{B_4'} we will then arrive at an imaginary term in $V_{\text{eff}}$ which we may freely tune via the gauge parameter $\beta$. 
This freedom to tune is of course suggestive of the conclusion that any such terms are non-physical.
Whilst the real-valued nature of $\beta$ prevents us from tuning any such terms exactly to zero, we may in practice tune such that they may no longer play any physical role; i.e. that 
	\begin{align}
		\text{Im}\left(\frac{24\pi^2}{\Lambda^2}V_{\text{eff}}\left(\sigma_c\right)\right)=2n\pi~, \quad
		n\in\mathbb{Z}~,
	\end{align}
for some non-trivial minimum $\sigma_c$ satisfying the self-consistency condition $V_{\text{eff}}\left(\sigma_c\right)=0$ outlined in the next section.
Given the inelegance of this approach however, we will not make use of it in what follows. \\

We will instead address this issue via an appeal to so-called `physical' gauges as outlined in \ref{sec:physical gauge}.
In the current context, this would amount to disregarding the components of the graviton corresponding to general coordinate transformations; i.e. the fields $\xi_\mu^T$ and $\chi$ in our notation.
One may note that it is the $`A3+\sqrt{A4}$' term which yields the (non-contributory) last element of \eqref{zetas}, and also by comparison with the pure GR case as in \eqref{grlimit} verify that it corresponds to the trace of the graviton $h$.
It is perhaps illustrative to then note that it is precisely $\xi_\mu^T$ and $\chi$  that correspond to the two remaining $\zeta'$ functions in \eqref{zetas}, indicating that any imaginary terms in this context must arise purely from gauge, rather than physical, degrees of freedom. 
Working in a physical gauge these $\zeta'$ functions would not be present, and reality of the action for negative $\Lambda_0$ would thus be assured.

\subsection{Effective potential}

Whilst, as discussed previously, we may tune $\beta$ to eliminate any imaginary terms, it is simpler in this context to utilise a physical gauge.
Having already derived the effective action $\Gamma$ in generality, it is straightforward to specialise to this gauge; we set $\left\{\alpha=\beta\to0\right\}$ and disregard the fields $\xi_\mu^T$ and $\chi$, along with the ghost fields that were originally introduced to cancel out their (gauge) degrees of freedom (which at any rate do not contribute for $\Lambda\to 0$).
The functional integral over these gauge degrees of freedom then yields an infinite constant prefactor of the volume of the diffeomorphism group, which is unimportant for our purposes.
As also noted in the previous section this process incurs extra zero modes, but whose contributions are subleading as $\Lambda\to0$ and so can also be neglecting in the following.
We find for $\left\{\alpha=\beta\to0\right\}$

	\begin{align}\label{Veff2}
		V_{\text{eff}}=f^2-\sigma_c^2
		+\frac{\kappa^4}{61952\pi^2}\Bigg(&
		16335 f^4
		-10890 \left(f^2-\sigma_c^2\right)^2 \ln \left(\frac{3 \kappa ^2 \left(f^2-\sigma_c^2\right)}{2 \mu ^2}\right)
		-32670 f^2 \sigma_c^2\nonumber\\
		&\;\;+61156 \sigma_c^4 \ln \left(\frac{\kappa ^2 \sigma_c^2}{3 \mu ^2}\right)
		-75399 \sigma_c^4
		+58564 \sigma_c^4 \ln\left(\frac{33}{2}\right)
		+2592 \sigma_c^4 \ln \left(\frac{54}{11}\right)\Bigg)~,
	\end{align}
where the limit $\Lambda \to 0$ is understood to have been taken.\\

We may firstly note the presence of the $\ln\left(\kappa^2\left(f^2-\sigma_c^2\right)\right)=\ln\left(-\Lambda_0\right)$ term in \eqref{Veff2}, which has the capability to destabilise the potential for $f^2<\sigma_c^2$.
In some sense, given the general incompatibility of supersymmetry with de Sitter space, this should perhaps be unsurprising: as our intention is to break local supersymmetry dynamically (i.e. via loop corrections), breaking it first at tree level via a positive cosmological constant $\Lambda_0$ renders the subsequent breaking via a dynamically generated $\sigma_c$ an impossibility.
As such, we must tune $f$ for a given value of $\mu$ to find self consistent minima $\sigma_c$ satisfying the condition $\sigma_c^2<f^2$ to ensure $\Lambda_0<0$ and thus a real $V_{\text{eff}}$.
If this condition is not met, $V_{\text{eff}}$ will contain an imaginary contribution
	\begin{align}
		\frac{45 i}{256\pi}\kappa^4\left(f^2-\sigma_c^2\right)^2.
	\end{align}

It is furthermore interesting to note that this problematic term arises precisely from the spin 2 part of the effective potential; absent these contributions, we find the potential
	\begin{align}	
		f^2-\sigma_c^2
		+\frac{\kappa^4}{30976\pi^2}\Bigg(\sigma_c^4 \left(30578 \ln \left(\frac{\kappa ^2 \sigma_c^2}{3 \mu ^2}\right)+29282 \ln
   \left(\frac{33}{2}\right)+1296 \ln \left(\frac{54}{11}\right)-45867\right)\Bigg)~,
	\end{align}
which is real for all $f$, $\sigma_c$.
We may compare this expression with the potential computed in \cite{smith2}, found in a similar context but via an expansion about a flat, rather than curved, spacetime (thus neglecting fluctuations of the metric field).
	\begin{align}\label{Vflat}
		V_{\text{flat}}=f^2-\sigma_c^2+\frac{\kappa^4}{4\pi^2}\left(11\frac{\sigma_c^2\Lambda'^2}{\kappa^2}+\sigma_c^4\left(\frac{121}{2}\ln\left(\frac{11\kappa^2\sigma_c^2}{\Lambda'^2}\right)-\frac{121}{4}\right)\right)~,
	\end{align}
where $\Lambda'$ is the UV cutoff implemented in \cite{smith2}.
The shapes of the potentials \eqref{Veff2} and \eqref{Vflat} are qualitatively similar, however, we find generically that \eqref{Veff2} leads to a larger dynamically generated gravitino mass.\\

\begin{figure}[h!!]
		\centering
		\includegraphics[width=0.45\textwidth]{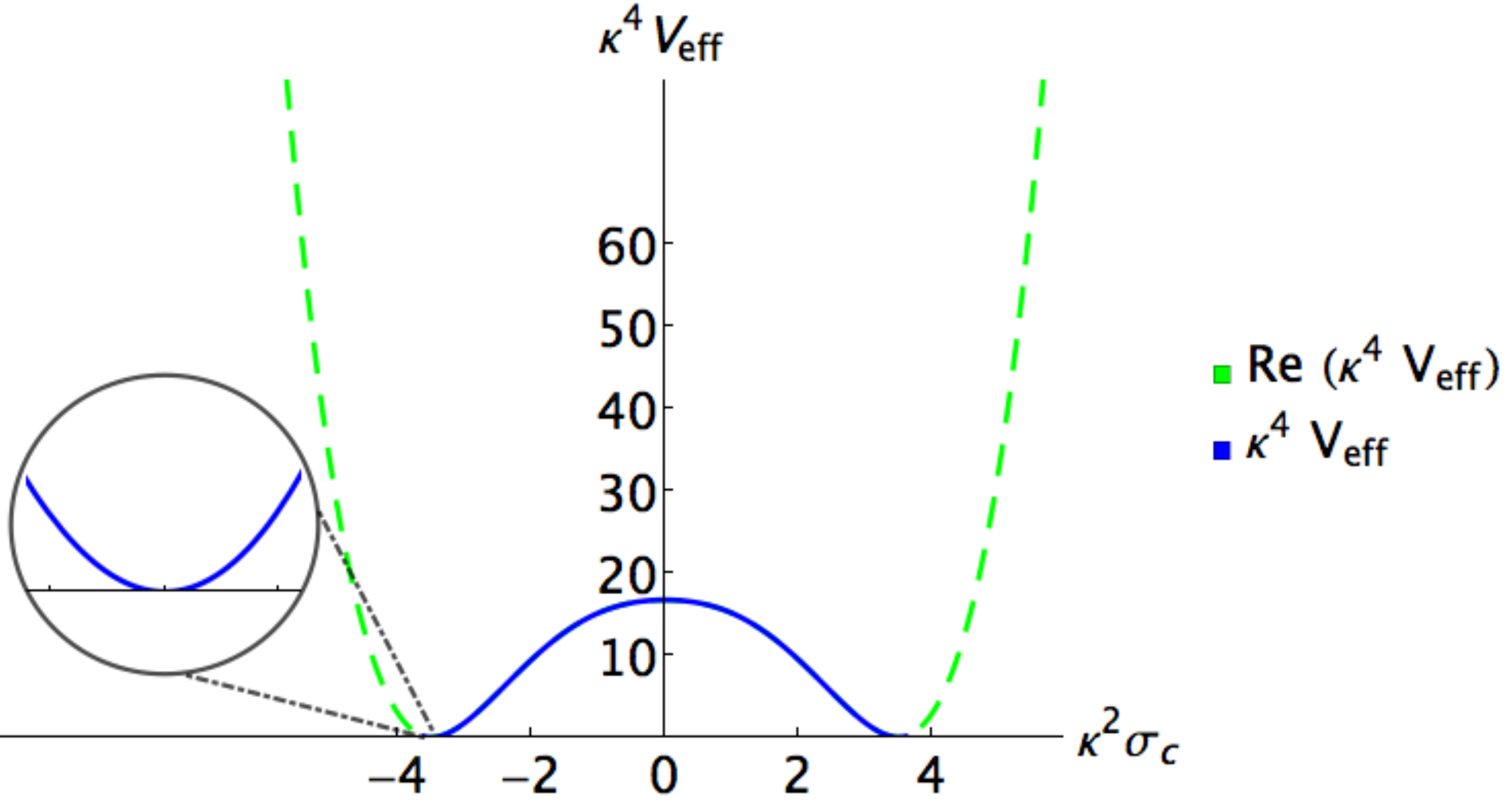}
		\caption{The effective potential \eqref{Veff2} for $\kappa\mu\simeq 3.87990$, $\kappa^2f\simeq 3.60559$, with non-trivial minima at $\kappa^2\sigma_c\simeq 3.50000$. We may observe firstly that $\Lambda_0<0$ and $V_{\text{eff}}=0$ at the minima, to respectively ensure reality and self-consistency with our previous limit $\Lambda\to0$, and secondly that the onset of imaginary terms (represented by the transition from blue (solid) to green (dashed) in the curve) occurs when $\Lambda_0$ changes sign.
		For higher values of the non-trivial minima, this transition occurs further and further away from the minimum, extending the range of the blue curve beyond the minimum at the cost of a higher dynamically generated mass.}
\label{fig:pot}	
\end{figure}

At this point we emphasise that the complexity of the potential beyond the self-consistent minima is unproblematic in the current context; the minima $\sigma_c$ is a constant which is self-consistently determined by the minimisation procedure, and in that sense is not a dynamical field. 
Quantum fluctuations about the minima, are represented by a space-time dependent condensate field $\sigma (x) = \sigma_c + 
\hbar {\tilde \sigma}(x)$, which however is massive, with mass of order of the gravitino field, 
as can be deduced by the (parabolic) shape of the effective potential around the non-trivial minima. 
Thus, such quantum fluctuations are suppressed and are not capable of destroying the stability of the broken phase minima. \\

A rigorous estimation of the mass of the gravitino condensate requires a computation of the wave-function renormalisation of the condensate field, following the steps outlined in \cite{eminfl}. 
In the present context, such an analysis has to be performed in de Sitter space prior to the limit $\Lambda \to 0$. 
Given that such estimates are of no interest to us at this stage, we will not perform these calculations here. 
The mass of the gravitino condensate, though, is essential when discussing the phenomenology of inflationary scenarios that such condensates may induce~\cite{eminfl}. 
This is postponed for a future publication. \\

It is straightforward to note (see fig.~\ref{fig:pot}) that the potential above has the correct shape to realise the super-Higgs effect, yielding (\emph{cf}. \eqref{ginomass}, in view of the normalisation of \cite{Fradkin:1983mq} \eqref{normmass}) a minimum dynamically generated gravitino mass of order
	\begin{equation}\label{gravinomass}
		m = \sqrt{\frac{11}{2}} \kappa\, \sigma_c = \sqrt{\frac{11}{16\pi}}\kappa^2\sigma_cM_{Pl}\simeq1.63730M_{Pl}\simeq 1.99899\times10^{19}~{\rm GeV}~.
	\end{equation} 
with a corresponding global supersymmetry breaking scale 
\begin{equation}\label{fscale}
\sqrt{f}\simeq 0.37876 M_{Pl}\simeq 4.62433\times10^{18}~{\rm GeV}~.
\end{equation} 	

Furthermore, we may isolate the fermionic and bosonic contributions at zero and one-loop order as
	\begin{align}
		V_{\text{eff}}=V_{B}^{(0)}+V_{B}^{(1)}+V_{F}^{(1)}
		=-\frac{\Lambda_0}{\kappa^2}+V_{B}^{(1)}+V_{F}^{(1)}
	\end{align}
so that we may identify
	\begin{align}\label{boson}
		V_{B}^{(1)}=\frac{45 \kappa ^4}{512 \pi^2}\left(f^2-\sigma_c ^2\right)^2 \left(3-2 \ln \left(\frac{3 \kappa ^2 \left(f^2-\sigma_c ^2\right)}{2 \mu ^2}\right)\right)~, 
		\end{align} 
		and 
		\begin{align}\label{fermion}
		V_{F}^{(1)}=\frac{\kappa ^4 \sigma_c^4}{30976 \pi ^2} \left(30578 \ln \left(\frac{\kappa^2 \sigma_c^2}{3 \mu ^2}\right)-45867+29282 \ln \left(\frac{33}{2}\right)+1296 \ln\left(\frac{54}{11}\right)\right)~.
	\end{align}
At the non-trivial minima given above we find $\kappa^4V_{F}^{(1)}\simeq-0.791357$, $\kappa^4V_{B}^{(1)}\simeq 0.0410402$, 
while the tree-level cosmological constant is negative (corresponding to anti de-Sitter space time), $\kappa^2\Lambda_0 \simeq-0.750279$. In this case, the one-loop cosmological constant (vacuum energy) vanishes, as a result of the stronger spin-2 contributions (positive sign) as compared with the (opposite sign) gravitino torsion terms.\\

The above-mentioned  dynamical gravitino mass can be modified by tuning the free quantities $f$ and $\mu$, with the above minimum value (\ref{gravinomass}), obtained when the non-trivial minima of the potential occur close to inconsistent regions of the parameter space, where $\Lambda_0$ changes from negative to positive, preventing the dynamical mass from being consistently decreased further.\\

	\begin{figure}[h!!]
		\centering
		\includegraphics[width=0.4\textwidth]{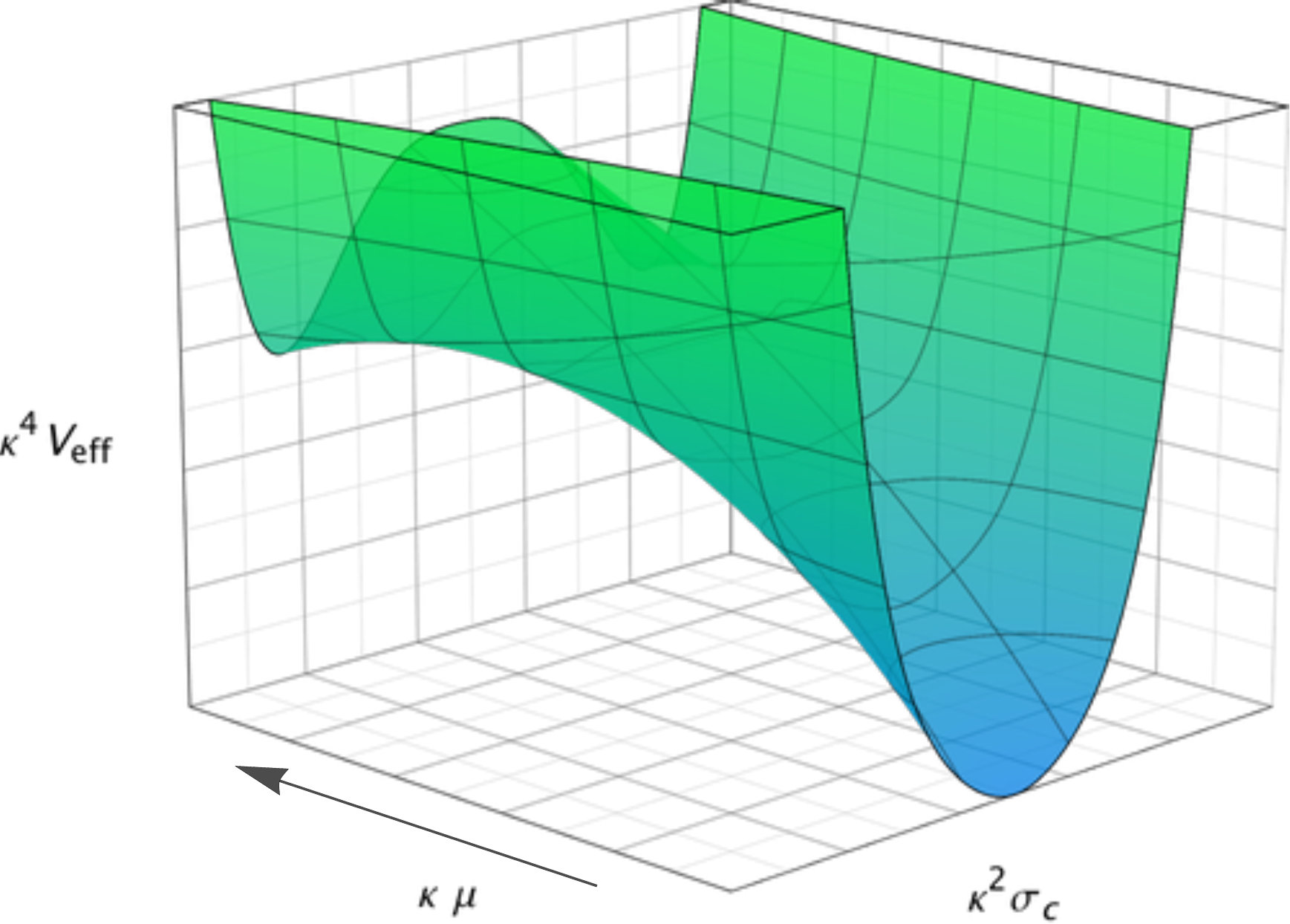}
		\includegraphics[width=0.4\textwidth]{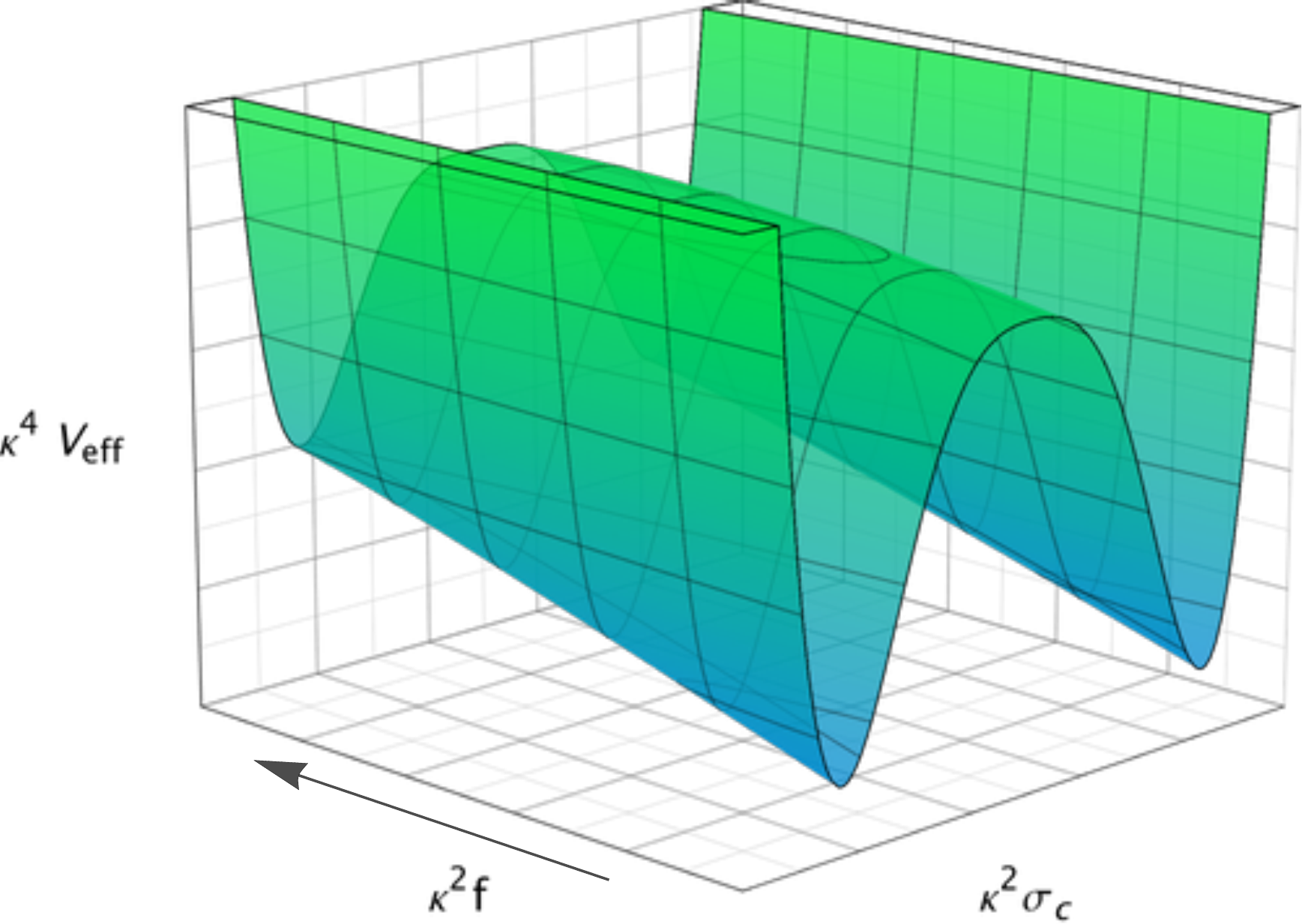}
		\caption{The effective potential \eqref{Veff2} showing schematically the effect of tuning $\mu$ and $f$, whilst holding, respectively, $f$ and $\mu$ fixed. The arrows in the respective axes 
		correspond to the direction of increasing $\mu$ and $f$. We may firstly note that as we flow from UV to IR (\emph{i.e}. in the direction of increasing $\mu$, \emph{cf}. discussion following Eq.~(\ref{mudef})) we obtain the correct double-well shape required for the super-Higgs effect, and secondly that tuning $f$ allows us to shift $V_{\text{eff}}$ and thus attain the correct vacuum structure (\emph{i.e.} non-trivial minima $\sigma_c$ such that $V_{\text{eff}}\left(\sigma_c\right)=0$).}
\label{fig:potmu}	
	\end{figure}

Moreover, as demonstrated in fig.~\ref{fig:potmu}, the shape of the effective potential changes, as one varies the 
(renormalisation) scale $\mu$ from Ultra-Violet (UV) to Infra-Red (IR) values (\emph{i.e.} flowing in the direction of increasing $\mu$), in such a way that the symmetry broken phase (double-well shaped potential) is reached in the IR. For comparison, we also mention that this feature also characterises the flat-space time effective potential of \cite{smith2}, as shown in \cite{eminfl}. This demonstrates that the dynamical generation of a gravitino mass is actually an IR phenomenon, in accordance with rather general features of dynamical mass in field theory. \\

\subsection{Conformal Supergravity models and phenomenologically realistic Gravitino masses \label{sec:confsugra}} 

Given the phenomenological unsuitability of the above (transplanckian) value
of the gravitino mass (\ref{gravinomass}) and the global supersymmetry breaking scale (\ref{fscale}),
of order of the reduced Planck mass, it is natural to seek ways for generating much lower gravitino masses, for global supersymmetry scales below the Grand Unification scale, thereby making contact with realistic phenomenology of such theories.
To this end, we shall consider an extension of the above analysis to a conformal ${\mathcal N}=1$ supergravity scenario, e.g. of the type considered in~\cite{confsugra}, which was used in \cite{eminfl} to discuss the role of gravitino condensates in providing inflationary scenarios compatible with the current astrophysical data~\cite{Planck}.\\

Although we shall postpone, for a forthcoming publication, a detailed analysis of dynamical breaking of local supersymmetry in such conformal supergravities, including their matter sectors (that notably can accommodate the next-to-minimal supersymmetric standard model), we may nevertheless discuss some important consequences of such models in yielding phenomenologically realistic values for the global supersymmetry breaking scale $f$ and the dynamical gravitino mass.\\

In this respect, we first notice that, in conformal supergravities~\cite{confsugra,emdyno} in the Jordan frame, there is a non-trivial coupling of a dilaton superfield to the gravitational Einstein term in the supersymmetric action. 
Passing to the Einstein-frame~\cite{emdyno}, where the (bosonic) Einstein-Hilbert scalar curvature term in the lagrangian assumes its canonically normalised form, the gravitational part of the action of ${\mathcal N}=1$ conformal supergravity, of relevance to our purposes here, is given by: 
\begin{eqnarray}\label{confsugra2}
\mathcal{L}^E  (e^E)^{-1} & = &  -\frac{1}{2\kappa^2} R^E(e^E)  - \frac{1}{2} \epsilon^{\mu\nu\rho\sigma} \overline{\psi '}_\mu  \gamma_5 \gamma_\nu D^E_\rho {\psi'}_\sigma -  e^{2\varphi}\,V^E
  \nonumber \\ &~&\,\,\,\, -\frac{11 \kappa^2}{16} e^{-2\varphi} \left( (\overline{\psi'}_\mu  {\psi'}^\mu )^2  - (\overline{\psi'}_\mu  \gamma_5 {\psi'}^\mu )^2 \right)
+ \frac{33}{64} \kappa^2 e^{-2\varphi} \, \left(\overline{\psi'}^\rho \gamma_5  \gamma_\mu {\psi'}_\rho  \right)^2  + \dots ~, 
\end{eqnarray}
where the superscript $E$ denotes quantities in the Einstein frame,
$\varphi$ is the (dimensionless) dilaton field, $V^E(\varphi, \dots)$ its potential, 
$\psi'_\mu$ denotes the canonically-normalized gravitino with a standard kinetic term as in $\mathcal{N}=1$ supergravity, and the 
$\dots$ in (\ref{confsugra2}) denote contributions from dilaton-derivative terms, dilatinos, as well as auxiliary, gauge-fixing ghost and matter fields, which will not play a role in our discussion here. 
\\

In general the form of the potential $V^E$ depends on the low-energy content of the action, and apart from the dilaton terms it may also contain terms that depend on matter multiplet fields that appear in next-to-minimal extensions of the standard model that can be accommodated in the low-energy limit of such frameworks.  For our purposes in this work we simply assume~\cite{eminfl} that, upon appropriately minimising the potential $V^E(\varphi, \dots)$, the dilaton field is stabilised to a (space-time) constant v.e.v. $\langle \varphi \rangle = \varphi_0$. In this way, we observe from (\ref{confsugra2}) that in conformal supergravities, the 
four-gravitino interaction terms carry an extra factor arising from the v.e.v. of the dilaton field, yielding a modified coupling in the gravitino self-interaction sector~\cite{eminfl}
	\begin{align}\label{kappatilde}
		\tilde\kappa=e^{-\varphi_0}\kappa~, \quad \langle \varphi \rangle = \varphi_0~.	
		\end{align}
The presence of two couplings, one (the standard gravitational one, $\kappa$) for the metric tensor interactions, and the other (${\tilde \kappa}$, \emph{cf}. Eq.~ (\ref{kappatilde}))	for the gravitino self-interaction terms, leads to the possibility of dynamical generation of much smaller gravitino masses  
in the instance of conformal supergravity models with~\cite{eminfl}  ${\tilde \kappa} \gg \kappa$
than in the simple ${\mathcal N}=1$ supergravity scenario discussed in previous sections. \\

It should be noted, however, that, as in the standard supergravity case, the tree-level cosmological constant at the broken symmetry minima (\ref{bare}) must be negative (anti-de-Sitter space time) in order to avoid imaginary parts in the potential, which still arise in the bosonic part of the potential when 
 $\sigma_c^2 > f^2 $. This is a notable difference from the flat space-time case of \cite{smith,smith2,eminfl}, where $\Lambda_0 > 0$ at the non-trivial minima, and thus global and local supersymmetries are already broken at tree level. We next notice that, in the conformal case the gravitino torsion parts of the effective potential dominate over the contributions due to the spin-2 graviton quantum fluctuations, and thus it is such torsion condensates which drive the one-loop cosmological constant to zero in this case. Since the situation is qualitatively similar to the flat space-time case of \cite{smith2}, we may say that in the case of conformal supergravity models with ${\tilde \kappa} \gg \kappa$, the gravitino torsion condensates ``effectively'' \emph{parallelize} the manifold, in the sense of the gravitino parts of the spin connection being responsible  for ``flattening'' the space-time~ 
\footnote{In this sense, this situation is somewhat reminiscent of the role of gravitino torsion contributions in the extra dimensional space ($S^7$, which is known to be a parallelisable manifold) of higher-dimensional (D=11) supergravities~\cite{englert}, which were argued to condense, 
cancelling any contributions from the metric in the corresponding components of the Christoffel symbol.
In this way, a vanishing cosmological constant arises~\cite{englert} in the four-dimensional space-time obtained after appropriate compactification.
However, our result on ``torsion-induced parallelism'' in the 4D conformal supergravity case does not constitute an exact mathematical statement, since it is only demonstrated at one loop. Unlike the higher-dimensional case of \cite{englert}, it has not been demonstrated that the gravitino torsion parts in the connection cancel out exactly any bosonic metric contributions to the
Christoffel symbols, thereby leading to a vanishing Riemann tensor for our four-dimensional space-time manifold.  
All we have shown here is that, at one loop order, the minimum of the effective potential vanishes, if there are Lorentz-invariant bilinear gravitino condensates, corresponding to a non zero gravitino mass. 
The resulting effective Einstein equations then, for vanishing cosmological constant, admit flat Minkowski solutions under the one-loop approximation. 
Hence there is no contradiction of the current results with any rigorous theorems on parallelism, known in the literature.}.\\
	
To see this, we first notice that in the conformal supergravity case, the contributions to the one-loop effective potential coming from the spin-2 (graviton) fluctuations are still given by (\ref{boson}), while the gravitino contributions are given by (\ref{fermion}) upon the replacement of $\kappa$ by ${\tilde \kappa}$:
\begin{align}\label{fermion2}
		V_{F}^{(1)}& =\frac{{\tilde \kappa} ^4 \sigma_c^4}{30976 \pi ^2} \left(30578 \ln \left(\frac{{\tilde \kappa}^2 \sigma_c^2}{3 \mu ^2}\right)-45867+29282 \ln \left(\frac{33}{2}\right)+1296 \ln\left(\frac{54}{11}\right)\right) \nonumber \\ 
& = \left(\frac{{\tilde \kappa}}{\kappa}\right)^4 \, \frac{\kappa^4 \sigma_c^4}{30976 \pi ^2} \left(30578 \ln \left(\left(\frac{{\tilde \kappa}}{\kappa}\right)^2\, \frac{\kappa^2 \sigma_c^2}{3 \mu ^2}\right)-45867+29282 \ln \left(\frac{33}{2}\right)+1296 \ln\left(\frac{54}{11}\right)\right)~.
	\end{align}

\begin{figure}[h!!]
		\centering
		\includegraphics[width=0.45\textwidth]{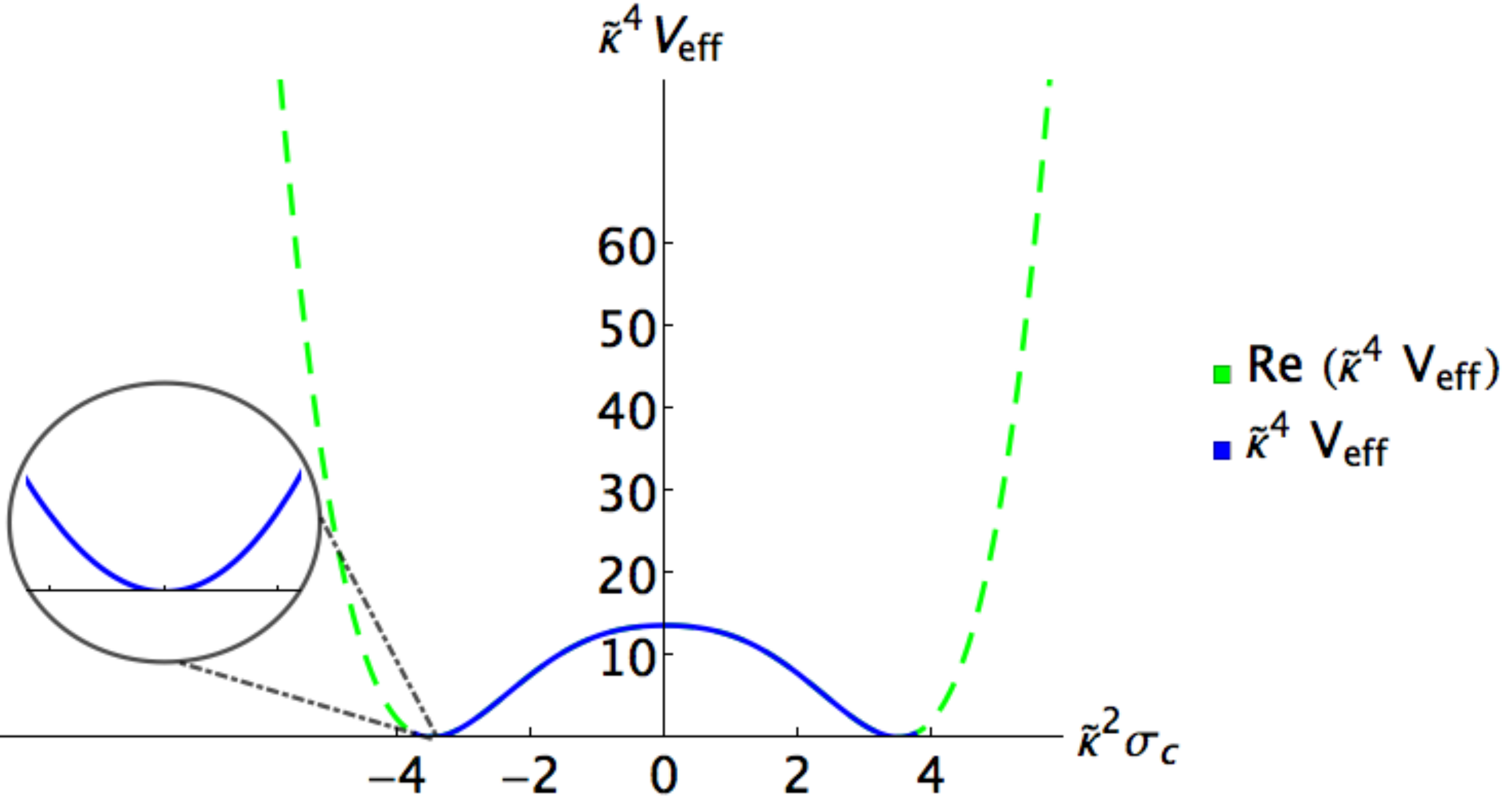}
		\caption{The effective potential  in the case of conformal supergravity models, with parameters 
$\tilde\kappa = 10^3 \kappa$, $\tilde\kappa^2\sigma_c \simeq 3.50000$, $\tilde\kappa^2f \simeq 3.69182$, $\tilde\kappa\mu \simeq 3.95668$. The flatness of the potential around the origin is pronounced  compared to the simple ${\mathcal N}=1$ supergravity case of fig.~\ref{fig:pot}, whilst the other features of the diagram remain unchanged.	}
\label{fig:confpot}	
\end{figure}

If we consider for concreteness the case $\tilde\kappa=10^3 \kappa$, which is a value dictated by the inflationary phenomenology of the model~\cite{eminfl}, we may find solutions with a vanishing one-loop effective potential at the non-trivial minima corresponding to (\emph{cf}. fig.~\ref{fig:confpot}):
\begin{eqnarray}\label{solutions}
{\tilde \kappa}^2 \, \sigma_c \simeq 3.50000~, \quad {\tilde \kappa}^2 \, f \simeq 3.69182~, \quad {\tilde \kappa}\, \mu \simeq 3.95668~, 
\end{eqnarray}
which lead to a global supersymmetry breaking scale 
	\begin{equation}\label{fscaleconf}
		\sqrt{f} \simeq 4.67933\times10^{15}~{\rm GeV}~,
	\end{equation}
 and dynamical gravitino mass 
 	\begin{equation}\label{gravinoconf}
 		m = \sqrt{\frac{11}{2}} \tilde\kappa\, \sigma_c 
		= \sqrt{\frac{11}{16\pi}}\left(\frac{\kappa}{\tilde\kappa}\right)\tilde\kappa^2\sigma_cM_{Pl}
		\simeq 1.99899\times10^{16}~{\rm GeV}~.
	\end{equation}
At the non-trivial minima we find $\tilde\kappa^4V_{F}^{(1)}\simeq-1.37957$, $\tilde\kappa^4V_{B}^{(1)}\simeq5.87744\times10^{-13}$, with tree-level cosmological constant $\tilde\kappa^2\Lambda_0 \simeq-1.37957$.
These values are phenomenologically realistic, thereby pointing towards the viability (from the point of view of producing realistic results of relevance to phenomenology) of the scenarios of dynamical breaking of local supersymmetry in conformal supergravity models. \\
 
An additional important feature is the pronounced flatness (near the origin, cf. fig.~\ref{fig:confpot}) of the one-loop effective potential for the gravitino condensate in such models, as compared to the pure supergravity  case of the previous sections. 
This leads~\cite{eminfl} to inflationary scenarios, with the massive gravitino condensate playing the role of the inflaton field, that are in agreement with the Planck satellite data on inflation~\cite{Planck}.  
The (slow-roll) inflationary phase in such approaches is identified with the rolling of the inflaton/gravitino-condensate field towards its non-trivial minimum on either side of the potential, with a model-dependent duration. 
The initial value of the inflaton depends on the initial conditions for inflation, whilst the exit from inflation coincides with the phase in which the field oscillates coherently near its non-trivial 
minimum. 
In that regime, as becomes evident from the shape of the double well potential, the inflaton condensate is massive (in the conformal model of \cite{eminfl} the inflaton has mass of the order of the grand unification scale). Reheating of the Universe then could conceivably arise from the decays of the condensate field upon coupling supergravity to matter. 
It would be interesting to explore such scenarios within our framework, in which the full quantum gravity corrections have been accounted for. This will constitute the topic of a future publication. \\

Finally, before closing this section we should also mention  an additional possibility that might characterise the conformal supergravity models of inflation, as envisaged in \cite{emdyno,eminfl}. This regards viewing the broken-supergravity-phase vacuum as being a \emph{metastable} one, with a lifetime longer than the age of the Universe. In such a case, the restriction on the value of the condensate field at the non-trivial minima $\sigma_c^2 < f^2$ could be lifted (thus implying local supersymmetry breaking already at tree-level), provided the conformally rescaled coupling ${\tilde \kappa}$ is sufficiently larger than the gravitational coupling $\kappa$, in order to ensure metastability. Detailed studies in this direction, making contact with realistic phenomenologies, will also be a topic for future work.
 
\section{Conclusions and Outlook \label{sec:concl} }

In this work we have revisited the issue of dynamical breaking of local supersymmetry (supergravity) without coupling to either gauge or matter sectors. 
We have examined for simplicity the specific case of $D=4$, ${\mathcal N}=1$ simple supergravity, although our considerations can be extended to more complicated and extended supergravities. 
We have constructed one-loop effective potentials, which exhibit a double-well (symmetric about the origin) shape and which vanish at the non-trivial minima.
In this way dynamical mass generation of the gravitino field can be understood most clearly. 
We have taken into account quantum fluctuations of the metric field by expanding about a Euclidean de Sitter background and taking the limit of a vanishing renormalised cosmological constant only at the end of the computations. \\

It is essential for our arguments that global supersymmetry is broken at a given scale. 
We have assumed F-type breaking for concreteness, which resulted in a positive value of the one-loop effective potential at the origin of the condensate field. 
This allows for the existence of a non-trivial minimum at which the one-loop effective potential vanishes, consistently with the assumed vanishing of the (renormalised) cosmological constant $\Lambda \to 0$. \\

We have demonstrated the existence of vacua at which the imaginary parts of the effective action were absent, contrary to previous claims in the literature~\cite{Buchbinder:1989gi}, thereby supporting the possibility of dynamical breaking of supergravity and the acquisition of a dynamical mass by the gravitino. The latter is found to be or order of the Planck mass 
in this simple supergravity framework, although this conclusion changes in more complicated settings, such as conformal supergravities~\cite{confsugra}, where phenomenologically realistic values are obtained for the dynamical gravitino mass and the global supersymmetry breaking scale. 
Our findings support therefore, at least qualitatively, the results of the flat-space analysis of \cite{smith,smith2} on dynamical breaking of supergravity. \\

An interesting feature of the double-well potentials that we have calculated in the broken phase of conformal supergravity models is their flatness about the origin, prompting one to consider small-field inflationary scenarios with the gravitino condensate playing the role of the inflaton field.
A preliminary study~\cite{eminfl} of such scenarios in the context of the flat-space potential of \cite{smith2} within a conformal supergravity model \cite{emdyno} has shown agreement of the inflationary cosmology with the current cosmological data. A complete analysis in this direction, within the context of conformal supergravitry models, where one-loop quantum fluctuations of the metric field are fully taken into account, 
as in the current article, is still pending, and will be the subject of a future work. \\

\section*{Acknowledgements} 

The work of N.H. is supported by a KCL GTA studentship, while that of 
N.E.M. is supported in part by the London Centre for Terauniverse Studies (LCTS), using funding from the European Research Council via the Advanced Investigator Grant 267352 and by STFC (UK) under the research grant ST/J002798/1.

\section*{Appendix}

In this Appendix we give details of some mathematical aspects and notions of our approach towards the construction of the one-loop effective action of ${\mathcal N}=1$ $D=4$ supergravity model. 
More precisely, we detail the use of the heat kernel in computing functional determinants, before specialising to the computation of the resultant zeta functions on $S^4$.
Finally, we demonstrate an asymptotic expansion which allows these zeta functions to be explicitly evaluated in the limit $\Lambda\to 0$.

\subsection*{The heat kernel}
Consider a second-order Laplace-type differential operator of the form $\Delta=-\Box+X$, for some constant $X$, defined on a smooth vector bundle over a compact, smooth $d$-dimensional Riemannian manifold without boundary. 
There exist a discrete number of eigenfunctions and corresponding eigenvalues of this operator, which may be decomposed spectrally into a complete orthonormal set of eigenfunctions $\phi_n$ with eigenvalues $\lambda_n$, of multiplicity $g_n$. \\

The determinant of this operator may be expressed
	\begin{align}
		\prod_{n}^{\infty}\lambda_{n}^{g_n}~,
	\end{align}
however as this obviously diverges, we shall instead define the zeta function
	\begin{align}\label{zeta}
		\zeta\left(z\right)\equiv\sum_{n}^{\infty}g_n\lambda_n^{-z}~,
	\end{align}
convergent for $\Re\left(z\right)>2$, which can in practice be extended via analytic continuation to a meromorphic function of $z$ over the entire complex plane.
It is important to note that $\zeta\left(z\right)$ is regular at $z=0$, yielding the derivative
	\begin{align}
		\zeta'\left(0\right)=-\sum_n g_n\ln\left(\lambda_n\right)~,
	\end{align}
		so that we may define 
	\begin{align} 
		\det\left(\Delta\right)\equiv\exp\left(-\frac{d}{dz}\zeta\left(z\right)\bigg|_{z=0}\right)~.
	\end{align}
Our task is then to compute the form of $\zeta'$ for a given operator.\\

A convenient way of encapsulating some of the behaviour of $\zeta$ is via the `trace over the heat kernel', defined thusly
	\begin{align}\label{a2}
		\Tr \;K\left(x,x,t,\Delta\right)\sim\sum_i b_i\left(x,\Delta\right)t^{\left(i-d\right)/2}~,
	\end{align}
valid for $t\to 0^{+}$, where $K$ satisfies the heat equation with boundary condition
	\begin{align}
		\frac{d}{dt}K\left(x,x',t,\Delta\right)+\Delta K\left(x,x',t,\Delta\right)=0~, \quad
		K\left(x,x',0,\Delta\right)=\delta\left(x,x'\right)~,
	\end{align}
and the $b_i\left(x,\Delta\right)$ are the heat kernel coefficients, which integrate to give spectral invariants of $\Delta$. 
Since the heat equation has the solution
	\begin{align} \label{a1}
		K\left(x,x',t,\Delta\right)=\sum_i\phi_i\left(x\right)\otimes\phi_i\left(x'\right)\exp\left(-t\lambda_i\right)~,
	\end{align}
we can trace over \eqref{a1} and integrate to see that (since the $\phi_i$ form an orthonormal basis)
	\begin{align}
		&\;\;\;\;\;\sum_i\exp\left(-t\lambda_i\right)\left(\phi_i,\phi_i\right)\left(x\right)\sim
		\sum_i b_i\left(x,\Delta\right)t^{\left(i-d\right)/2}\\
		&\label{a3}\Rightarrow \sum_i\exp\left(-t\lambda_i\right)\sim\sum_i  t^{\left(i-d\right)/2}\int\sqrt{g} \;b_i\left(x,\Delta\right)d^{d}x
		\equiv\sum_{i=0}  t^{\left(i-d\right)/2}B_i\left(\Delta\right)~.
	\end{align}
Finally, we note that $\zeta\left(z\right)$ is related to \eqref{a3} via a Mellin transform
	\begin{align}\label{Mellin}
		\zeta\left(z\right)
		=\frac{1}{\Gamma\left(z\right)}\int_{\epsilon\to 0^+}^{\infty}t^{z-1}\sum_i\exp\left(-t\lambda_i\right)dt
		=\frac{1}{\Gamma\left(z\right)}\int_{\epsilon\to 0^+}^{\infty}\sum_i t^{\left(i+2z-d-2\right)/2}B_i\left(\Delta\right)dt~,
	\end{align} 
so that for $t\to 0^+$ we may expand the sum (thus preserving only the first few terms) and extract information about $\zeta$ as necessary.\\

Taking the divergent parts given by the trace over the heat kernel, in conjunction with the known expression for the finite parts of zeta regularised determinants \cite{Hawking:1976ja}, we can then find for $d=4$
	\begin{align}
		\ln\det\left(\frac{\Delta_s}{\mu^2}\right)=
		-\frac{1}{2}B_0 L^4-\frac{1}{2}B_2 L^2
		-B_4\left(\ln\left(\frac{L^2}{\mu^2}\right)
		-\gamma\right)
		-B_4\ln\left(\frac{\rho^2}{\mu^2}\right)
		+\zeta_s'\left(0\right)~,
	\end{align}
where we have a cut-off $\epsilon=\left(\mu^2/L^2\right)\to0$, and $\gamma$ is the Euler-Mascheroni constant. 
The mass dimension of $\mu$ and $L$ is one, and it is important to note at this point that as \eqref{Mellin} can be though of as a proper time integral, $\epsilon\to0$ is a short time (and thus high energy) cutoff.
Flowing from the UV to IR therefore corresponds to the direction of increasing $\mu^2$.\\

Computing the form of the $b_i$ is straightforward in practice as their general forms are known \cite{Fradkin:1983mq}.
In $d=4$ we have
	\begin{align}
		B_p=\frac{1}{16\pi^2}\int \overline{b}_p\sqrt{g}d^4x~, \quad
		\overline{b}_0=\Tr\mathbb{1}~,\quad
		\overline{b}_2=\Tr\left(\frac{1}{6}R-X\right)~,\quad
		b_i=\left(4\pi\right)^2\overline b_i~,
	\end{align}
where the trace is performed in the space of fields.	
As the heat kernel coefficients are straightforward to find (for low $d$), our problem of evaluating functional determinants is now reduced to computing the form of $\zeta\left(0\right)$ and $\zeta'\left(0\right)$ for a given background. 

\subsection*{Zeta functions}	
As $\zeta$ is defined by the eigenvalues and their degeneracies for a given operator, we must work in a framework where these quantities are known.
This is achieved in practice by specialising to `differentially constrained' operators; those corresponding to irreducible representations of the background isometry group.
For $SO\left(5\right)$ these representations can be labelled by $(n,s)$ (for a positive integer $n$ and a spin $s$) with corresponding eigenvalues $\lambda\left(n,s\right)$ and degeneracies then given by the dimension of the representation, 
	\begin{align}
		\lambda\left(n,s\right)=n\left(n+3\right)+2ns+\frac{X}{\rho}~,
		\quad g_n=\dim\left(n,s\right)=\frac{1}{6}\left(2s+1\right)\left(n-s+1\right)\left(n+s+2\right)\left(2n+3\right)~.
	\end{align}
For the various representations we therefore have \cite{Fradkin:1983mq} (where we have rescaled $\lambda_n=\rho\overline\lambda_n,$ $X=\rho\overline X$, for $\rho=\Lambda/3$ on $S^4$)\\

\noindent {\bf Spin 0: $\left(n,0\right)$}
	\begin{align}
		\overline{\lambda}_n=n^2+3n+\overline{X}, \quad
		g_n=\frac{1}{6}(n+1)(n+2)(2n+3)
	\end{align}
{\bf Spin 1: $\left(n,1\right)$}	
	\begin{align}
		\overline{\lambda}_n=n^2+3n-1+\overline{X}, \quad
		g_n=\frac{1}{2}(n+3)(2n+3)
	\end{align}
{\bf Spin 2: $\left(n,2\right)$}
	\begin{align}
		\overline{\lambda}_n=n^2+3n-2+\overline{X}, \quad
		g_n=\frac{5}{6}(n-1)(n+4)(2n+3)
	\end{align}	
{\bf Spin 1/2: $\left(n\pm\frac{1}{2},\frac{1}{2}\right)$}	
Where both representations have the same spectra, for (n-1/2,1/2) we have
	\begin{align}
		\overline{\lambda}_n=(n+1)^2+\overline{X}, \quad
		g_n=\frac{2}{3}n(n+1)(n+2),
	\end{align}
which is to be doubled.	\\

\noindent {\bf Spin 3/2: $\left(n\pm\frac{1}{2},\frac{3}{2}\right)$}	
Where both representations again have the same spectra, for (n-1/2,3/2) we have
	\begin{align}
		\overline{\lambda}_n=(n+1)^2+\overline{X}, \quad
		g_n=\frac{4}{3}(n-1)(n+1)(n+3)
	\end{align}
which is to be doubled.	
Spinor representations may be incorporated into this framework via `squaring' the corresponding first-order operators to yield those of second-order.\\

With this in mind, our prior expression for $\zeta$ \eqref{zeta} can now be re-expressed more concretely as
	\begin{align}
		\zeta_s\left(z,X\right)
		&=\sum_{n=0}^{\infty}g_n\lambda_n^{-z}
		=\frac{\left(2s+1\right)}{6}\sum_{n=0}^{\infty}\frac{\left(n-s+1\right)\left(n+s+2\right)\left(2n+3\right)}{\left(n\left(n+3\right)+2ns+\frac{X}{\rho}\right)^{z}}
		=\frac{1}{3}\left(2s+1\right)F\left(z,2s+1,\left(s+\frac{1}{2}\right)^2,b_s\left(X\right)\right)~,
	\end{align}
where (following the conventions of \cite{Fradkin:1983mq}) we have defined	
	\begin{align}
		F\left(z,k,a,b\right)\equiv\sum_{v=\frac{1}{2}k+1}^{\infty}\frac{v(v^2-a)}{(v^2-b)^z}~,
	\end{align}
	\begin{align}
		b_0\left(X\right)=\frac{9}{4}-\frac{X}{\rho}~,\; 
		b_{1/2}\left(X\right)=-\frac{X}{\rho}~,\; 
		b_1\left(X\right)=\frac{13}{4}-\frac{X}{\rho}~,\; 
		b_{3/2}\left(X\right)=-\frac{X}{\rho}~,\; 
		b_2\left(X\right)=\frac{17}{4}-\frac{X}{\rho}~,
	\end{align}
where $\rho=\Lambda/3$, and we note that our sum starts from the minimal $v$ such that $g_n>0$, and therefore all possible negative and zero modes are included. \\
	
Following the appendix of \cite{Christensen:1979iy}, it is firstly straightforward to show that
	\begin{align}
		F\left(0,k,a,b\right)=\frac{1}{4}b\left(b-2a\right)+\frac{1}{24}a\left(3k^2+6k+2\right)-\frac{1}{64}k^2\left(k+2\right)^2+\frac{1}{120}~.
	\end{align}
In line with this derivation we can compute $F\left(1,k,a,b\right)$ via a similar argument, making use of the identity \cite{Allen:1983dg}
	\begin{align}
		\sum_{n=0}^\infty\sum_{m=1}^\infty\frac{b^{n}}{\left(m+\frac{k}{2}+1\right)^{-\left(2n+1\right)}}
		=-\frac{1}{2}\left(\digamma\left(\frac{k}{2}+1+\sqrt{b}\right)+\digamma\left(\frac{k}{2}+1-\sqrt{b}\right)\right)~,
	\end{align}
where $\digamma$ is the digamma function. 
We find
	\begin{align}
		F\left(1,k,a,b\right)=\frac{1}{2}b-\frac{1}{12}
		-\frac{1}{8}k\left(k+2\right)
		-\frac{1}{2}\left(b-a\right)\digamma\left(\frac{k}{2}+1\pm\sqrt{b}\right)~.
	\end{align}
We may then note that
	\begin{align}
		\frac{d}{db}F'\left(0,k,a,b\right)=F\left(1,k,a,b\right)~,
	\end{align}
so that we may integrate both sides to arrive at
	\begin{align}\label{poly}
		F'\left(0,k,a,b\right)=\frac{1}{4}b^2-\frac{1}{12}b-\frac{1}{8}b k\left(k+2\right)
		-\frac{1}{2}\int_0^b\left(y-a\right)\digamma\left(\frac{k}{2}+1\pm\sqrt{y}\right) dy+C~,
	\end{align}
where $C$ is a (real) constant of integration
	\begin{align}
		C=F'\left(0,k,a,0\right)=2\zeta_R'\left(-3,\frac{1}{2}k+1\right)-2a\zeta_R'\left(-1,\frac{1}{2}k+1\right)~, \quad 
		\zeta_R\left(z,q\right)=\sum_{n=q}^\infty n^{-z}~.
	\end{align}
	
For large $b$ we can explicitly evaluate the integral above by shifting the measure via $y\to y^2$ and inserting the asymptotic expansion 
	\begin{align}
		\digamma\left(x\right)
		=\ln\left(x\right)-\frac{1}{2x}-\sum_{n=1}^{\infty}\frac{B_{2n}}{2nx^{2n}}
		=\ln\left(x\right)-\frac{1}{2x}-\frac{1}{12x^2}+\dots,
	\end{align}
where $B_n$ is the $n$th Bernoulli number. 
Integrating term by term, only the leading order remains relevant and we find
	\begin{align}
		F'\left(0,k,a,b\right)
		\sim\frac{1}{4}b^2-\int_0^{\sqrt{b}} y^3\log\left(\frac{k}{2}+1\pm y\right) dy
		\sim\frac{b^2}{8} \left(3-2\log \left(-b\right)\right)~.
	\end{align}

Combing these elements, we thus find that for small $\Lambda$
	\begin{align}
		\zeta_s\left(0,X\right)\sim\frac{6s+3}{4}\frac{X^2}{\Lambda^2}~,\quad
		\zeta'_s\left(0,X\right)\sim\frac{6s+3}{4}\frac{X^2}{\Lambda^2} \left(\frac{3}{2}-\ln\left(\frac{3X}{\Lambda}\right)\right)~.
	\end{align}
From this result, it is interesting to note that in the limit $\Lambda\to0$, the presence or absence of imaginary terms in the effective potential is a straightforward consequence of the sign of $X$, the argument of the functional determinant being evaluated. 
This may be therefore leveraged to explicitly investigate the possibility of instabilities in the effective potential, along with the associated question of whether or not any such apparent instabilities are, in fact, one-loop artefacts which may then be disregarded.
These issues are of course central to the topic of this paper.

\end{document}